\documentclass[10pt,aps,prb,amsmath,amssymb,twocolumn]{revtex4-2}
\usepackage{blindtext}
\usepackage{graphicx}
\usepackage{subfigure}
\usepackage{adjustbox}
\usepackage{bm}
\usepackage{color}
\usepackage{braket}
\usepackage{standalone}
\usepackage{multirow}
\usepackage{tikz}
\usepackage{mathrsfs}
\usepackage{dsfont}
\usepackage{mathtools}
\usepackage{comment}
\usepackage{mathtools}
\usepackage{float}
\usepackage[colorlinks,bookmarks=true,citecolor=blue,linkcolor=blue,urlcolor=blue]{hyperref}
\usepackage{cleveref}

\DeclarePairedDelimiter\abs{\lvert}{\rvert}%

\begin{document}
\title{Quantum Sensing with Driven-Dissipative Su-Schrieffer-Heeger 
Lattices}
\author{Oscar Arandes and Emil J. Bergholtz}
\affiliation{Department of Physics, Stockholm University, AlbaNova University Center, 10691 Stockholm, Sweden}
\date{\today}

\begin{abstract}
The remarkable sensitivity of non-Hermitian systems has been extensively studied and stimulated ideas about developing new types of sensors. In this paper, we examine a chain of parametrically driven coupled resonators governed by the squeezed Su-Schrieffer-Heeger model. We emphasize the qualitative difference in sensor performance between configurations depending on bulk topology and boundary modes, specifically for detecting both on-site and non-Hermitian skin effect perturbations. Our analysis goes beyond the scenario of infinitesimal perturbations, extending to arbitrary perturbation strengths beyond the linear response regime. We stress the importance of optimizing the system's parameters to achieve quantum enhancement while avoiding fine-tuned regimes that could limit the practical applicability of this system for real-world quantum sensing.
\end{abstract}

\maketitle

\section{Introduction}
\label{sec:intro}
Quantum sensing is a promising concept, not only for its potential to deepen our understanding of physics, but also for its future technological applications. In this context, non-Hermitian (NH) physics, which has received a surge of attention in recent years, may provide valuable insights. Although non-Hermiticity has long been studied in the context of both classical and open quantum systems, a significant amount of research over the past decade has introduced a new perspective on band topology, extending it beyond the conventional Hermitian framework in condensed matter physics~\cite{NHbook,NHreview,ShenZhenFu2018,GongAshida2018,KaShUeSa2019,KönigYang2023,YangLiKönig2024}. The field has rapidly evolved, revealing a variety of phenomena, including spectral singularities known as exceptional points (EP)~\cite{MiriAlu2019}, exponential localization of energy eigenstates referred to as the non-Hermitian skin effect (NHSE)~\cite{Lee2016,YaoWang2018,KunstEdvardsson2018,Okuma2020,GhatakCoulais2020,ZhangLuChen2022,LinTaiLiLee}, and scale-free localized states~\cite{GuoHuChen2023,LiWangSongWang2023,MoligniniArandesBergholtz2023,XieLiangMaDuPeng}, among others. The unique sensitivity of NH systems exhibiting the skin effect to perturbations~\cite{KunstEdvardsson2018,Xiong2018,Edvardsson2022,Trefethen}, sparked discussions on their potential for enhancing response in the field of sensing~\cite{NHsensor}.

Earlier proposals aimed to exploit the square-root dependence of frequency splitting on small perturbations at EPs to develop a new kind of classical sensors with an enhanced response~\cite{WiersigEP2016,ZhongRenKhajavikhan2019}, contrasting with the linear dependence observed in Hermitian systems. Theoretical developments later expanded these ideas to the quantum regime~\cite{ZhangSweeneyHsuYang,LiuLiDuWenYang2024}. Experimental validation was reported in various classical platforms, including coupled optical ring resonators~\cite{ChenÖzdemirZhao,HodaeiHassanWittek,WangHouLu} and electromechanical accelerometers~\cite{KononchukCaiEllis}. However, shortly after the initial theoretical proposals for EP sensors, other works cast doubt on whether the claimed enhancements were as significant as anticipated. Consideration of the role of eigenstates, rather than just eigenvalues, and a proper accounting of system noise, began to reveal that these sensors may not surpass their Hermitian counterparts as initially expected~\cite{Langbein,ChenJinLiu,DugganMan,LoughlinSudhir,WangLaiYuan}. The use of different formalisms to model NH phenomena, each valid across different regimes, makes it difficult to determine whether EP sensors will prove advantageous in the future~\cite{WiersigEP2020}. 

The exploration of non-Hermiticity's potential to enhance sensing has since gone beyond EPs. Some studies have proposed that non-reciprocity can enhance sensing performance even away from an EP~\cite{LauClerk,PengZhuZhang}, while others argue that non-Hermitian sensors provide no fundamental advantage~\cite{DingWangChen}. 

Extended NH systems featuring the NHSE exhibit exponential amplification (and deamplification) of the wave function~\cite{WanjuraBrunelli2020} as well as a diverging Petermann factor in the absence of exceptional points~\cite{Schomerus2020}. 
Concretely, by manipulating the boundary conditions, these phenomena can be exploited for a qualitatively new type of sensors, initially proposed in the classical regime~\cite{NHsensor} and later extended to the quantum regime~\cite{KochBudich2022,WangZhuWangZhangChong2022}. In addition, it has led to innovative proposals for NH cooling mechanisms~\cite{XuDelicWangCapellaroLi2024}. NHSE sensors have already been performed on classical experimental platforms, such as electrical circuits~\cite{ZhangZhouWangPan,NHTopOhm}.

In the full quantum regime, NH dynamics have naturally emerged in Hermitian quadratic bosonic Hamiltonians that do not conserve particle number~\cite{YuXinWang}, such as in arrays of coupled resonators with nonlinearities introduced by parametric driving. Amplifying behaviors were first observed in a 1D lattice implementing the bosonic Kitaev chain (BKC)~\cite{McDonald2018}. Specifically, it was shown that the system's dynamics were closely related to one of the canonical examples of NH tight-binding models, namely the Hatano-Nelson model~\cite{HatanoNelson,HatanoNelson2,HatanoNelson3}. Experimental realizations of the BKC and Hatano-Nelson dynamics have been achieved in optomechanical platforms~\cite{SlimWanjura}, quantum simulations~\cite{Busnaina} and time-multiplexed photonic resonator networks~\cite{LeefmansDutt2022,PartoLeefmansWilliamsMarandi2023}. 

The first theoretical proposal using the BKC for quantum sensing evaluated the sensor's ability to detect both a perturbation coupling the ends of an odd chain via the NHSE effect and a local on-site perturbation~\cite{McDonald2020}. While the original study concluded that the NHSE was not effective for sensing, as the signal amplification was counterbalanced by a proportional increase in the number of photons traveling through the coupled resonators, subsequent research argued that this conclusion overlooked proper optimization of system parameters, claiming that NHSE can, in fact, be a valuable resource for quantum sensing~\cite{BaoQiDong2022}. 

Recent efforts have explored further generalizations of these works by extending the system dynamics from the Hatano-Nelson model to the two-band tight-binding NH Su-Schrieffer-Heeger (SSH) model~\cite{SSH1,SSH2}. The SSH model is notable for its sublattice structure, with two sites per unit cell. Importantly, a sensor based on a chain of driven, coupled bosonic modes, whose dynamics are described by an NH SSH model with an even number of sites has been recently discussed~\cite{ZhangChen}. Although less conventional in the literature, using the SSH model with a \textit{broken} last unit cell (resulting in an \textit{odd} number of sites) exhibits a different boundary mode behavior and hence different response.

In this paper, we study a chain of driven, coupled bosonic modes comparing results for a system with a \textit{broken} final unit cell with the more standard SSH setup. The system's dynamics are governed by a NH SSH model 
and non-reciprocal coupling at \textit{every} site, offering a broader generalization of previous studies~\cite{ZhangChen}. Our analysis focuses on a regime where no growing modes are present, ensuring the stability of the sensor setup. We begin by considering a sensor designed to detect infinitesimal local on-site perturbations and NHSE perturbations. The discussion includes the differences between our setup and configurations with an even number of sites, where the last unit cell is unbroken, identifying the parameter regimes where the odd configuration outperforms the even one. For the NHSE perturbation, we demonstrate that in a model with an odd number of sites, a \textit{single} drive is sufficient to probe the system. In contrast, for a model with an even number of sites, the signal for the NHSE perturbation is zero, requiring at least two drives to generate a finite response~\cite{ZhangChen}. We also extend our analysis beyond linear order, relaxing the requirement for infinitesimal perturbations. This leads to results that generalize previous studies on the BKC~\cite{McDonald2020}. Notably, we demonstrate that the NHSE can serve as a promising platform for quantum sensing, even beyond the linear response regime. 

\section{Model}
\label{sec:model}
\begin{figure}[!t]
    {\includegraphics[width=0.5\textwidth]{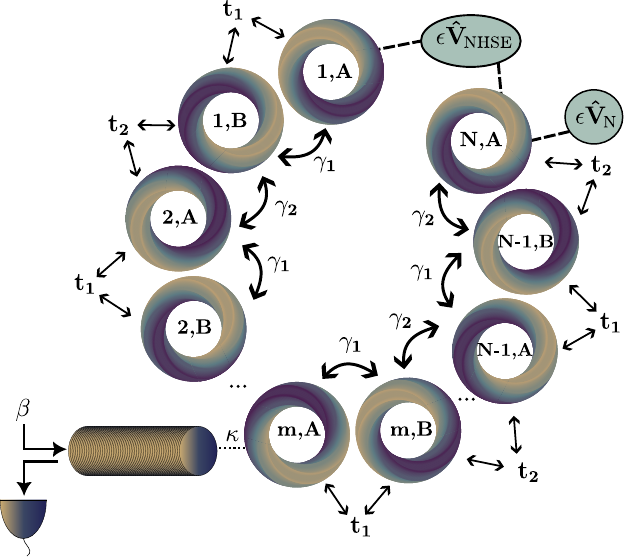}} 
    \caption{Chain of $2N-1$ coupled modes in ring geometry representing the Hamiltonian in Eq.~\eqref{eq:HamSystem}. Each unit cell comprises two sites, labeled as $A$ and $B$. The intracell couplings and parametric driving are denoted by $\gamma_1$ and $t_1$, respectively, while the intercell couplings and and parametric driving are represented by $\gamma_2$ and $t_2$. A coherent drive $\beta$ is injected through a waveguide at the $m,A$ site, where a homodyne measurement is also performed. Two types of perturbations are considered: an on-site potential applied to the last \textit{broken} unit cell and a perturbation coupling the chain's endpoints (NHSE perturbation).}
     \label{fig:system}
\end{figure}

The model we study is the so-called squeezed SSH model~\cite{WanLu}. We consider a chain of $N$ bosonic modes, each with sublattice structures A and B in the unit cell. We consider a chain where the last mode features a \textit{broken} unit cell, consisting of $2N-1$ (odd) number of sites, as depicted in FIG.~\ref{fig:system}. In the rotating frame where the parametric driving frequency matches the resonance frequency of the modes, the (Hermitian) Hamiltonian reads 
\begin{equation}
\begin{alignedat}{2}
\hat{H}_S = &\sum_{n=1}^{N-1} [&&i \, t_1 \,(\hat{a}_{n,A}\hat{a}_{n,B}-\hat{a}_{n,A}^\dagger\hat{a}_{n,B}^\dagger) \\[-0.7em]
& &&+i \, \gamma_1 \,(\hat{a}_{n,A}\hat{a}_{n,B}^\dagger-\hat{a}_{n,A}^\dagger\hat{a}_{n,B})] \\[-0.7em]
+&\sum_{n=2}^{N} [&&i \,t_2 \,(\hat{a}_{n,A}\hat{a}_{n-1,B}-\hat{a}_{n,A}^\dagger\hat{a}_{n-1,B}^\dagger) \\[-0.7em]
& &&-i \, \gamma_2 \,(\hat{a}_{n,A}\hat{a}_{n-1,B}^\dagger-\hat{a}_{n,A}^\dagger\hat{a}_{n-1,B})] ,
\end{alignedat}
\label{eq:HamSystem}
\end{equation}
where $t_1,t_2 \in \mathbb{R}$ are the strengths of the intra- and intercell squeezing terms, respectively, and the intra- and intercell nearest-neighbor hopping strengths are denoted by $\gamma_1,\gamma_2>0$. We take $\gamma_1>t_1$ and $\gamma_2>t_2$ to ensure dynamical stability (see Appendix~\ref{app:stability}). The model with $\gamma_2=0$ was recently studied for an even number of sites (i.e., when the chain is unbroken)~\cite{ZhangChen}. 

Physical realizations of this model can be achieved across various photonic and phononic experimental platforms. The essential feature of parametric driving in our model relies on the presence of nonlinearities. These can be realized through three-wave mixing processes, such as those in quantum superconducting circuits setups~\cite{AbdoKamal,Fitzpatrick,Frattini,KrantzKjaergaard,WangCurtis} or by exploiting the nonlinear optomechanical interaction in photonic crystals~\cite{AspelmeyerMarquardt,BrooksBotter,Safavi,PinoSlim}. Intrinsic intracavity nonlinearities of ring resonators could offer an alternative platform~\cite{McDonald2018,MittalGoldschmidt}.

\begin{figure}[!t]
    {\includegraphics[width=0.5\textwidth]{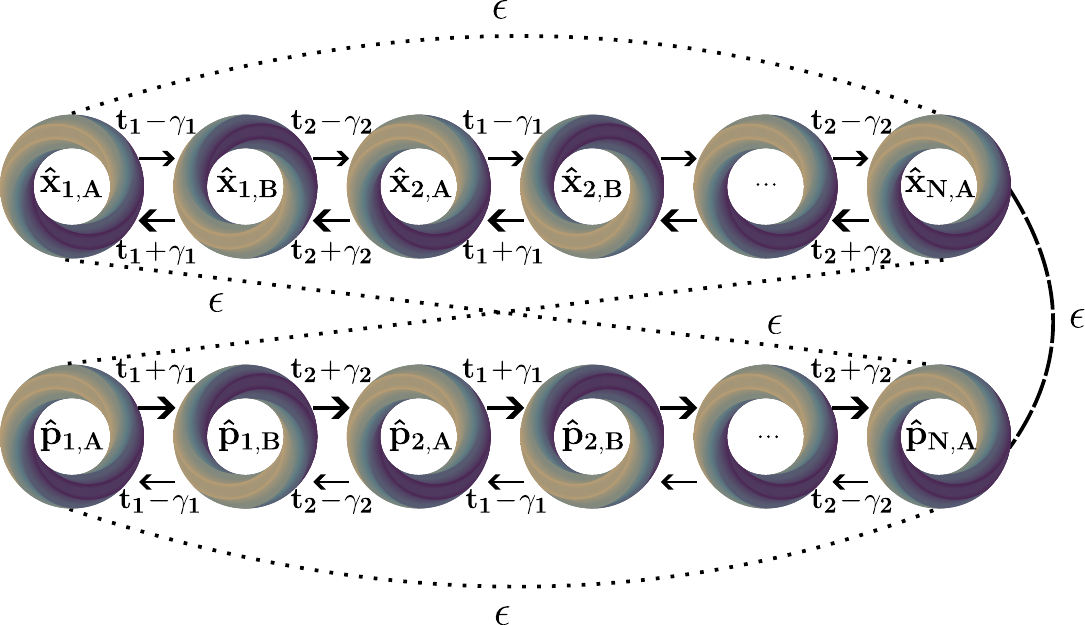}} 
    \caption{Dynamics of a chain with $2N-1$ coupled modes described by the equations of motion~\eqref{eq:eomquad}. Each unit cell consists of two sites, labeled as $A$ and $B$. In the absence of any perturbation, the dynamics correspond to two uncoupled non-Hermitian SSH chains in the $\hat{x}$ and $\hat{p}$ quadrature basis. The on-site perturbation, indicated by a dashed line, introduces coupling between the two chains at their last site. Meanwhile, the NHSE perturbation, shown with a dotted line, couples both the endpoints of each individual chain and links the $\hat{x}$-chain with the $\hat{p}$-chain as shown in the figure.}
     \label{fig:system_quad}
\end{figure}

The dynamics of Hamiltonian in Eq.~\eqref{eq:HamSystem} are governed through the Heisenberg equations of motion
\begin{equation}
    \begin{split}
        \dot{\hat{a}}_{n,A} &= -t_1 \, \hat{a}_{n,B}^\dagger - \gamma_1 \, \hat{a}_{n,B} - t_2 \, \hat{a}_{n-1,B}^\dagger + \gamma_2 \, \hat{a}_{n-1,B} ,\\
        \dot{\hat{a}}_{n,B} &= -t_1 \, \hat{a}_{n,A}^\dagger + \gamma_1 \, \hat{a}_{n,A} - t_2 \, \hat{a}_{n+1,A}^\dagger - \gamma_2 \, \hat{a}_{n+1,A} .
    \end{split}
\end{equation}
The connection with non-Hermitian dynamics becomes apparent when using the local canonical quadrature basis $\hat{a}_n = \frac{1}{\sqrt{2}}(\hat{x}+i\,\hat{p})$ in which the Hamiltonian is expressed as
\begin{equation}
   \begin{split}
    \hat{H}_S =  \sum_{n=1}^{N-1} [(\gamma_1\!-\!t_1) \, \hat{x}_{n,A}\,\hat{p}_{n,B} 
    - (\gamma_1\!+\!t_1) \, \hat{x}_{n,B}\,\hat{p}_{n,A}] \\
    -  \sum_{n=2}^{N} [ (\gamma_2\!+\!t_2) \, \hat{x}_{n,A}\,\hat{p}_{n-1,B} 
    + (\gamma_2\!-\!t_2) \,  \hat{x}_{n-1,B}\,\hat{p}_{n,A}] ,
   \end{split}
\label{HamSystemQuad}
\end{equation}
and the equations of motion given by
\begin{equation}
    \begin{split}
        \dot{\hat{x}}_{n,A} &= -(\gamma_1\!+\!t_1) \, \hat{x}_{n,B} + (\gamma_2\!-\!t_2) \, \hat{x}_{n-1,B} ,\\
        \dot{\hat{x}}_{n,B} &= +(\gamma_1\!-\!t_1) \, \hat{x}_{n,A} - (\gamma_2\!+\!t_2) \, \hat{x}_{n+1,A} ,\\
        \dot{\hat{p}}_{n,A} &= -(\gamma_1\!-\!t_1) \, \hat{p}_{n,B} + (\gamma_2\!+\!t_2) \, \hat{p}_{n-1,B} ,\\
        \dot{\hat{p}}_{n,B} &= +(\gamma_1\!+\!t_1) \, \hat{p}_{n,A} - (\gamma_2\!-\!t_2) \, \hat{p}_{n+1,A} .
    \end{split}
    \label{eq:eomquad}
\end{equation}
In this basis, it is now transparent that the dynamics of each $\hat{x}$ and $\hat{p}$ quadratures are decoupled and governed by two non-Hermitian SSH chains with opposite chirality (see FIG.~\ref{fig:system_quad}). For open boundary conditions, the SSH model exhibits a resonant zero mode for all parameter values, while for the analogous chain with an unbroken last unit cell, two localized zero modes are predicted by the biorthogonal polarization~\cite{KunstEdvardsson2018,EdvardssonKunstYoshida2020} (or, in presence of chiral symmetry, equivalently, the non-Bloch winding number~\cite{YaoWang2018}) appearing at
\begin{equation}
    t_1 = \pm \sqrt{\gamma_1^2+t_2^2-\gamma_2^2} \,, \qquad t_1 = \pm \sqrt{\gamma_1^2-t_2^2+\gamma_2^2} \,,
\end{equation}
and remained localized at the end of the chain as long as the condition
\begin{equation}
    \left|\frac{\gamma_2\!+\!t_2}{\gamma_1\!-\!t_1} \right| > \left|\frac{\gamma_1\!+\!t_1}{\gamma_2\!-\!t_2} \right| ,
    \label{eq:evenmodes}
\end{equation}
is satisfied.

\section{Parameter estimation. Non-Hermitian sensor}
\label{sec:NHsensor}
In order to use our setup for parameter estimation, we consider two types of perturbations. The first is described by adding the term $\epsilon \,\hat{V}_{N} = \epsilon \, \hat{a}_{N,A}^\dagger \hat{a}_{N,A}$ to the Hamiltonian in Eq.~\eqref{eq:HamSystem}, where $\epsilon$ is the parameter we aim to detect when it acquires a non-zero value. This perturbation corresponds to an on-site perturbation at the last site of the chain~\cite{McDonald2020,BaoQiDong2022}.

The second perturbation, denoted by $\epsilon \, \hat{V}_{\text{NHSE}} = \epsilon \, (e^{i\phi}\,\hat{a}_{1,A} \hat{a}_{N,A}^\dagger + e^{-i\phi}\,\hat{a}_{1,A}^\dagger \hat{a}_{N,A})$, represents a perturbation connecting the first and last sites of the chain. It is termed the NHSE perturbation due to its association with the extreme sensitivity exhibited by non-Hermitian systems when changing its boundary conditions~\cite{McDonald2020,BaoQiDong2022,NHsensor,KochBudich2022,RafiUlIslam2024}. We take $\epsilon$ to have units of frequency and $\hat{V}$ to be dimensionless.

To probe the system, we inject a coherent drive into one of the modes in the A sublattice (hereafter labeled $m$) via an input-output waveguide. We employ a homodyne-detection measurement scheme connected to the same $m$-mode to extract the signal from the system. The Heisenberg equations of motion for the system modes and the waveguide can be then derived using the standard input-output formalism~\cite{GardinerCollett,ClerkDevoret}, yielding
\begin{equation}
    \begin{split}
    \dot{\hat{a}}_{n,A} = &-t_1 \, \hat{a}_{n,B}^\dagger - \gamma_1 \, \hat{a}_{n,B} - t_2 \, \hat{a}_{n-1,B}^\dagger + \gamma_2 \, \hat{a}_{n-1,B} \\
    &-i \,\epsilon \,[\hat{a}_{n,A},\hat{V}]  \\
    &-\frac{\kappa}{2}\,\hat{a}_{m,A}\,\delta_{n m} - \sqrt{\kappa}\, (\beta + \hat{B}^{\text{in}})\,\delta_{n m} ,\\
        \dot{\hat{a}}_{n,B} = &-t_1 \, \hat{a}_{n,A}^\dagger + \gamma_1 \, \hat{a}_{n,A} - t_2 \, \hat{a}_{n+1,A}^\dagger - \gamma_2 \, \hat{a}_{n+1,A}\\
        &-i \,\epsilon \,[\hat{a}_{n,B},\hat{V}] .
    \end{split}
    \label{EomInOut}
\end{equation}
From the perspective of the $m$-mode in the cavity, the coupling with the waveguide induces damping with a decay rate $\kappa>0$. Additionally, there will be an effect due to the injection of photons from the classical drive $\beta = \abs{\beta} e^{i\theta}$ and the quantum noise entering through the waveguide $\hat{B}^{\text{in}}$ (for details see Ref.~\onlinecite{LauClerk,BaoQiDong2022}). Markovian dynamics are already implicit in Eq.~\eqref{EomInOut}, where $\hat{B}^{\text{in}}$ is assumed to satisfy the standard quantum Gaussian white noise correlation functions
\begin{equation}
    \begin{split}
    \langle \hat{B}^{\text{in}}(t)\hat{B}^{\text{in} \, \dagger}(t') \rangle &= (\bar{n}_{\text{th}}  \! +\! 1) \,\delta(t-t') ,\\
    \langle \hat{B}^{\text{in}\,\dagger}(t)\hat{B}^{\text{in}}(t') \rangle &= \bar{n}_{\text{th}} \, \delta(t-t') ,\\
    \langle \hat{B}^{\text{in}}(t)\hat{B}^{\text{in}}(t') \rangle &= 0  ,
    \end{split}
\end{equation}
where $\bar{n}_{\text{th}}$ is the thermal occupation number for the input field $\hat{B}^{\text{in}}(t)$.

The outgoing light in the waveguide in the $m$-mode is given by the standard input-output relation 
\begin{equation}
    \hat{B}^{\text{out}}(t) = \beta + \hat{B}^{\text{in}}(t) + \sqrt{\kappa} \, \hat{a}_m(t) ,
    \label{Bout}
\end{equation}
which we integrate over a long time period $\tau$ to obtain the temporal mode of the output field
\begin{equation}
    \hat{\mathcal{B}}_\tau (N) = \frac{1}{\sqrt{\tau}} \int_0^\tau dt \, \hat{B}^{\text{out}}(t) ,
    \label{TempOutMode}
\end{equation}
where we have made explicit the dependence with the system size $N$. The normalization constant in the definition in Eq.~\eqref{TempOutMode} ensures the canonical commutation relation $[\hat{\mathcal{B}}_\tau (N), \hat{\mathcal{B}}_\tau^\dagger (N)]=1$.

In \textit{local} quantum estimation theory~\cite{Paris}, we aim to find the optimal measure scheme that minimizes the variance $\delta \epsilon$ of the estimator at a fixed value of the parameter
\begin{equation}
    \delta \epsilon = \sqrt{\langle (\epsilon_{\text{est}}-\epsilon)^2\rangle} .
\end{equation}
It can be shown that any quantum measurement is bounded through the quantum Cramér-Rao inequality, which for an unbiased estimator reads as
\begin{equation}
    \delta \epsilon \geq \frac{1}{\sqrt{M \, \mathcal{F}}} ,
\end{equation}
where $M$ is the number of measurements and $\mathcal{F}$ is the so-called quantum Fisher information (QFI). The optimal measurement scheme for our linear Gaussian system that maximizes the QFI is given by a standard homodyne measurement in the large drive limit $(\abs{\beta}\gg 1)$~\cite{BianchiBraunstein,PinelJian}. The homodyne current is given by
\begin{equation}
    \hat{\mathcal{M}}_\tau (N) = \frac{1}{\sqrt{2}} \left( e^{-i\varphi}\hat{\mathcal{B}}_\tau (N) + e^{i\varphi} \hat{\mathcal{B}}_\tau^\dagger (N)   \right) ,
    \label{HomoCurrent}
\end{equation}
where $\varphi$ is the measurement angle in phase space that allows us to control which quadrature we want to measure (for $\varphi = 0$ and $\varphi = \frac{\pi}{2}$ we would measure the $\hat{x}$- and $\hat{p}$- quadratures, respectively).

In the large drive limit, it can be shown that the QFI can be directly related to the signal-to-noise ratio~\cite{McDonald2020,BaoQiDong2022,LauClerk} 
\begin{equation}
    \text{SNR}_\tau (N,\epsilon) \equiv \frac{\mathcal{S}_\tau(N,\epsilon)}{\mathcal{N}_\tau(N,\epsilon)} ,
\end{equation}
where the signal and noise power are given in terms of  $\hat{\mathcal{M}}_\tau (N)$ as
\begin{equation}
\mathcal{S}_\tau(N,\epsilon) = \abs{\langle \hat{\mathcal{M}}_\tau (N) \rangle_\epsilon -  \langle \hat{\mathcal{M}}_\tau (N) \rangle_0  }^2  , 
\label{SignalDef}
\end{equation}
and
\begin{equation}
\mathcal{N}_\tau(N,\epsilon) = \langle \hat{\mathcal{M}}_\tau^2 (N) \rangle_\epsilon - \langle \hat{\mathcal{M}}_\tau (N) \rangle_\epsilon^2 ,
\label{NoiseDef}
\end{equation}
where the expectation values are with respect to the steady state of Eq.~\eqref{EomInOut}.

In order to study the enhancement of the SNR as a function of the system size $N$, it is customary to take as a metric the SNR average with the total number of photons to eliminate the trivial enhancement of the SNR due to a larger number of photons when increasing the system size~\cite{McDonald2020,BaoQiDong2022,LauClerk}. We take
\begin{equation}
    \overline{\text{SNR}}_\tau (N,\epsilon) \, \equiv \, \frac{\text{SNR}_\tau(N,\epsilon)}{\bar{n}_{\text{tot}}(N,\epsilon)} ,
\end{equation}
with
\begin{equation}
   \begin{split}
    \bar{n}_{\text{tot}}(N,\epsilon) &= \sum_n \!\Big(\langle \hat{a}^\dagger_{n,A} \hat{a}_{n,A} \rangle_\epsilon + \langle \hat{a}_{n,B}^\dagger \hat{a}_{n,B}\rangle_\epsilon \Big)\\
    &\approx \sum_n  \!\Big( \langle \hat{a}^\dagger_{n,A}\rangle_\epsilon \langle \hat{a}_{n,A} \rangle_\epsilon + \langle \hat{a}_{n,B}^\dagger \rangle_\epsilon \langle \hat{a}_{n,B}\rangle_\epsilon \Big), \!\!
    \end{split}
    \label{PhotonsDef}
\end{equation} 
where the last approximation holds in the large-drive limit, where the coherent photons dominate over the photons from the amplified vacuum fluctuations. 

\section{SNR in the linear response regime}
\label{sec:SNRlinear}
\begin{figure}[htbp]
\includegraphics[width=\linewidth,trim={0 10 0 10},clip]{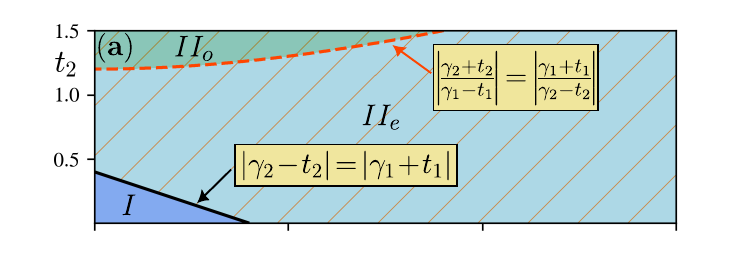}
\includegraphics[width=\linewidth,trim={0 0 0 10},clip]{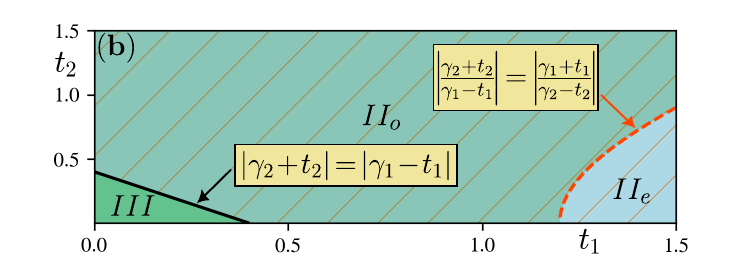}
\caption{
Phase diagram according to the response of the model in Eq.~\eqref{eq:HamSystem} as a function of parameters $t_1>0$ and $t_2>0$. For the {\it on-site perturbation}: blue areas indicate regions where the even chain outperforms the odd chain, while green areas represent the opposite. Regimes $I$ and $II$ are depicted in the upper figure, while regimes $II$ and $III$ are shown in the lower figure. Regime $II$ is divided by an orange dashed line, marking the transition at which the edge modes in the even chain emerge. In region $II_e$ the edge modes of the even chain are present (even chain having a steeper exponential scaling), whereas in region $II_o$, they are absent (odd chain demonstrating better performance). The striped lines highlight where the {\it NHSE perturbation} in the odd chain leads to a (non-fine-tuned) enhancement. Note that the even chain with the NHSE perturbation is not shown, as it results in a zero response. We fix the parameters as follows: (a)  $\gamma_1=1.6$ and $\gamma_2=2$. (b) $\gamma_1=2$ and $\gamma_2=1.6$. In all figures we set $\theta=\frac{\pi}{4}$, $\varphi=0$ and $\phi=\frac{\pi}{2}$. For the on-site perturbation we consider $m=1$, while for the NHSE perturbation we assume a linear scaling with the system size $m=\alpha N$.}
\label{fig:phasediagram}
\end{figure}

We begin by treating the perturbation strength $\epsilon$ to be infinitesimally small, allowing for an initial approximation of the sensor's behavior across different parameters. We first consider the on-site perturbation $\hat{V}_{N} = \hat{a}_{N,A}^\dagger \hat{a}_{N,A}$. When the conditions $\abs{\gamma_2\!+\!t_2}>\abs{\gamma_1\!-\!t_1}$ and $\abs{\gamma_2\!-\!t_2}>\abs{\gamma_1\!+\!t_1}$ (hereafter labeled as \textit{regime I}) are met, the odd chain exhibits exponential attenuation across most of the parameter range, regardless of the homodyne angle $\varphi$ or the drive phase $\theta$. In contrast, the even chain shows the opposite behavior, displaying exponential amplification~\cite{ZhangChen}. This amplification occurs in the regime where the condition in Eq.~\eqref{eq:evenmodes} is trivially satisfied, i.e., the even chain hosts a pair of zero modes localized at the chain ends. On the contrary, for $\abs{\gamma_2\!+\!t_2}<\abs{\gamma_1\!-\!t_1}$ and $\abs{\gamma_2\!-\!t_2}<\abs{\gamma_1\!+\!t_1}$ (\textit{regime III}) we observe the reverse scenario: the odd chain undergoes exponential amplification, while the even chain experiences exponential attenuation. Significantly, it is in this regime where the even chain has no zeros modes. In the intermediate regime where $\abs{\gamma_2\!+\!t_2}>\abs{\gamma_1\!-\!t_1}$ and $\abs{\gamma_2\!-\!t_2}<\abs{\gamma_1\!+\!t_1}$ (\textit{regime II}) the behavior of even and odd chain are similar, but with different rates of exponential enhancement. We find that the even chain outperforms the odd chain whenever two localized edge modes are present (see FIG.~\ref{fig:phasediagram}). Finally, we note that setting $\varphi=\theta=\{0,\frac{\pi}{2}\}$ results in a null SNR for both the odd and even chains.

Next, we fix the site we drive to $m=1$ and measure the $x$-quadrature by setting $\varphi=0$. We set the drive phase to  $\theta=\frac{\pi}{2}$ for simplicity, noting that a drive with $\theta=\frac{\pi}{4}$  (see FIG.~\ref{fig:SNRfirst}) would result in nearly identical behavior for the chosen parameters (see Appendix~\ref{app:DynamicsFirstOnsitePert} for a discussion for other parameter values). We focus on the regimes where the odd chain exhibits amplification (i.e., \textit{regimes II} and \textit{III}). The signal defined in Eq.~\eqref{SignalDef} for the on-site perturbation is 
\begin{equation}
    \mathcal{S}_\tau(N,\epsilon) = 32 \, \tau \left(\frac{\epsilon}{\kappa}\right)^2 \abs{\beta}^2 \left( \frac{\gamma_1\!+\!t_1}{\gamma_2\!-\!t_2}\right)^{\!\!4(N-1)}  .
    \label{eq:signalOnsite}
\end{equation}
Crucially, since the perturbation strength is infinitesimal, we only need to consider the zeroth order contribution from the noise in Eq.~\eqref{NoiseDef}, which does not undergo any amplification. It is given by 
\begin{equation}
     \mathcal{N}_\tau(N,0) = \bar{n}_{\text{th}} \!+\! \frac{1}{2} .
     \label{eq:noisezero}
\end{equation}
In the following, we set $\bar{n}_{\text{th}}=0$ without loss of generality. Finally, since the number of photons expression in Eq.~\eqref{PhotonsDef} scales in the long $N$ limit as
\begin{equation}
    \bar{n}_{\text{tot}}(N,0) \approx \frac{4}{\kappa} \abs{\beta}^2 \,\left( \frac{\gamma_1\!+\!t_1}{\gamma_2\!-\!t_2} \right)^{\!\!2(N-1)} ,
\end{equation}
we can derive an exponential enhancement
\begin{equation}
    \text{SNR}_\tau(N) \approx  16 \, \tau \, \kappa \,\bar{n}_{\text{tot}}  \left(\frac{\epsilon}{\kappa}\right)^2 \left( \frac{\gamma_1\!+\!t_1}{\gamma_2\!-\!t_2} \right)^{\!\!2(N-1)}  ,
\end{equation}
that persists even when the number of photons is kept constant. In \textit{regime II}, we can immediately recover the behavior of the SNR for the BKC by setting $t_2=t_1, \gamma_2=\gamma_1$ and $\bar{N}=2N-1$, where $\bar{N}$ is the number of unit cells for the BKC~\cite{McDonald2020}. 

We next consider the NHSE perturbation given by $\hat{V}_{\text{NHSE}} = e^{i\phi}\,\hat{a}_{1,A} \hat{a}_{N,A}^\dagger + e^{-i\phi}\,\hat{a}_{1,A}^\dagger \hat{a}_{N,A}$. In \textit{regime I}, the system always exhibits exponential attenuation, regardless the choice of $\varphi$ and $\theta$. In  \textit{regime II}, exponential amplification can be found, for instance,
 with $\overline{\text{SNR}}_\tau (N) \!\propto\! \left( \frac{\gamma_1+t_1}{\gamma_2-t_2}\right)^{2N}$ for $m=1, \theta = 0$ and $\varphi \in [0,\frac{\pi}{2})$. This represents a direct generalization of analogous amplifying paths observed in the BKS model, as reported in previous studies~\cite{BaoQiDong2022}. Notably, some of the amplification or attenuation in this regime depends sensitively on the parameters $\varphi$ and $\theta$ (see Appendix~\ref{app:DynamicsFirstNHSEPert}). Deviations from certain precise parameter values show that the amplification affecting the signal also appears in the average photon number, implying no true enhancement for the NHSE perturbation~\cite{McDonald2020}. In \textit{regime III}, a similar \textit{fine-tuned} amplification and attenuation behaviors are observed, ultimately confirming the absence of any non-trivial enhancement in this regime as well. 

 Fortunately, this is not the end of the story for the NHSE sensing. Rather than fixing the drive position $m$ and studying how the $\overline{\text{SNR}}_\tau$ varies with $N$, let us instead assume a linear scaling of the drive position with the system size, such that $m=\alpha\,N$ for $\frac{1}{N} \leq \alpha \leq 1$~\cite{ZhangChen}. Remarkably, we find that the exponential enhancement is recovered. Importantly, this enhancement does not require fine-tuning of the drive's phase or the homodyne angle, as it holds for a broad range of phases, specifically for $\theta \in (0,\frac{\pi}{2})$ and $\phi \in (0,\frac{\pi}{2}]$. Additionally, for fixed parameters $t_1$, $t_2$, $\gamma_1$, $\gamma_2$, there exists an optimal value $\alpha^*$ where the $\overline{\text{SNR}}_\tau$ shows the steepest scaling with the system size $N$, given by
 \begin{equation}
    \left( \frac{\gamma_2\!+\!t_2}{\gamma_1\!-\!t_1} \right)^{\!\!(\alpha^* N-1)} \approx \left( \frac{\gamma_1\!+\!t_1}{\gamma_2\!-\!t_2} \right)^{\!\!N(1-\alpha^*)} .
    \label{condCaseII}
\end{equation}
It is worth noting that for a given optimal value of $\alpha^*$, the chain must have a minimum size $N_{\text{min}}$ to achieve exponential enhancement, as determined by the condition $\alpha^*(t_1,t_2,\gamma_1,\gamma_2)\, N_{\text{min}}\geq 1$.
We set the parameters $\varphi=0$, $\theta=\frac{\pi}{4}$ and $\phi=\frac{\pi}{2}$, for simplicity. In \textit{regime II}, the signal for the NHSE perturbation can be found to be 
\begin{equation}
    \!\!\!\mathcal{S}_\tau(N,\epsilon) \!=\! 16  \tau \!\left(\frac{\epsilon}{\kappa}\right)^{\!2} \!\!\abs{\beta}^2 \!\! \left(\!\frac{\gamma_2\!+\!t_2}{\gamma_1\!-\!t_1}\!\right)^{\!\!2(\alpha N-1)} \!\!\!\left(\!\frac{\gamma_1\!+\!t_1}{\gamma_2\!-\!t_2}\!\right)^{\!\!2N(1-\alpha)} \!\!\!\!\!\!\!\!\! . \!\!\!
\end{equation}
At zeroth order, both the noise and the photon number are independent of the perturbation. The noise is given by Eq.~\eqref{eq:noisezero} and the expression for the number of photons, based on the parameters set for the NHSE perturbation, is given as
 \begin{equation}
   \!\!\!\bar{n}_{\text{tot}}(N,0)\! =  \!\frac{2 \abs{\beta}^2}{\kappa}  \!\!\left[\! \!\left( \!\frac{ \gamma_2 \! + \! t_2 }{\gamma_1\!-\!t_1} \!\right)^{\!\!2(\alpha N -1)} \! \! \! \! + \! \!\left(\! \frac{\gamma_1\!+\!t_1}{\gamma_2\!-\!t_2} \!\right)^{\!\!2N(1-\alpha)} \! \right] \!\! . \!\!\!
\end{equation} 
Two distinct regimes for the scaling of the SNR are identified, separated by the condition in Eq.~\eqref{condCaseII}. For the parameter region where $\abs{\frac{\gamma_1+t_1}{\gamma_2-t_2}}^{(1-\alpha)N}>\abs{\frac{\gamma_2+t_2}{\gamma_1-t_1}}^{(\alpha N-1)}$ (i.e., $\alpha\neq 1$) the SNR has the following expression
\begin{equation}
    \text{SNR}_\tau (N) = 16 \, \kappa\, \tau \, \bar{n}_{\text{tot}} \left( \frac{\epsilon}{\kappa} \right)^2  \left(\frac{\gamma_2\!+\!t_2}{\gamma_1\!-\!t_1}\right)^{\!\!2(\alpha N -1)}  ,
\end{equation}
while for $\abs{\frac{\gamma_1+t_1}{\gamma_2-t_2}}^{(1-\alpha)N}<\abs{\frac{\gamma_2+t_2}{\gamma_1-t_1}}^{(\alpha N-1)}$ (i.e., $\alpha\neq \frac{1}{N}$) we get
\begin{equation}
    \text{SNR}_\tau (N) = 16\,\kappa\, \tau \, \bar{n}_{\text{tot}} \left( \frac{\epsilon}{\kappa} \right)^2  \left( \frac{\gamma_1\!+\!t_1}{\gamma_2\!-\!t_2} \right)^{\!\!2N(1-\alpha)} .
\end{equation}
Thus, the NHSE perturbation remains useful for sensing applications, provided the system parameters are appropriately chosen. Importantly, this does not require a fine-tuned setup. The \textit{odd} chain offers a significant advantage compared to the even chain: in the even chain, the signal is found to be zero at first order~\cite{ZhangChen}. Achieving a non-zero and comparable scaling in the even chain would require at least two drives, while our system achieves it with \textit{only} a single drive. This makes the odd chain a simpler and more effective platform than the even chain for enhancing the response as the system size $N$ grows. 

\begin{figure}[htbp]
    \centering
    {\includegraphics[width=0.45\textwidth]{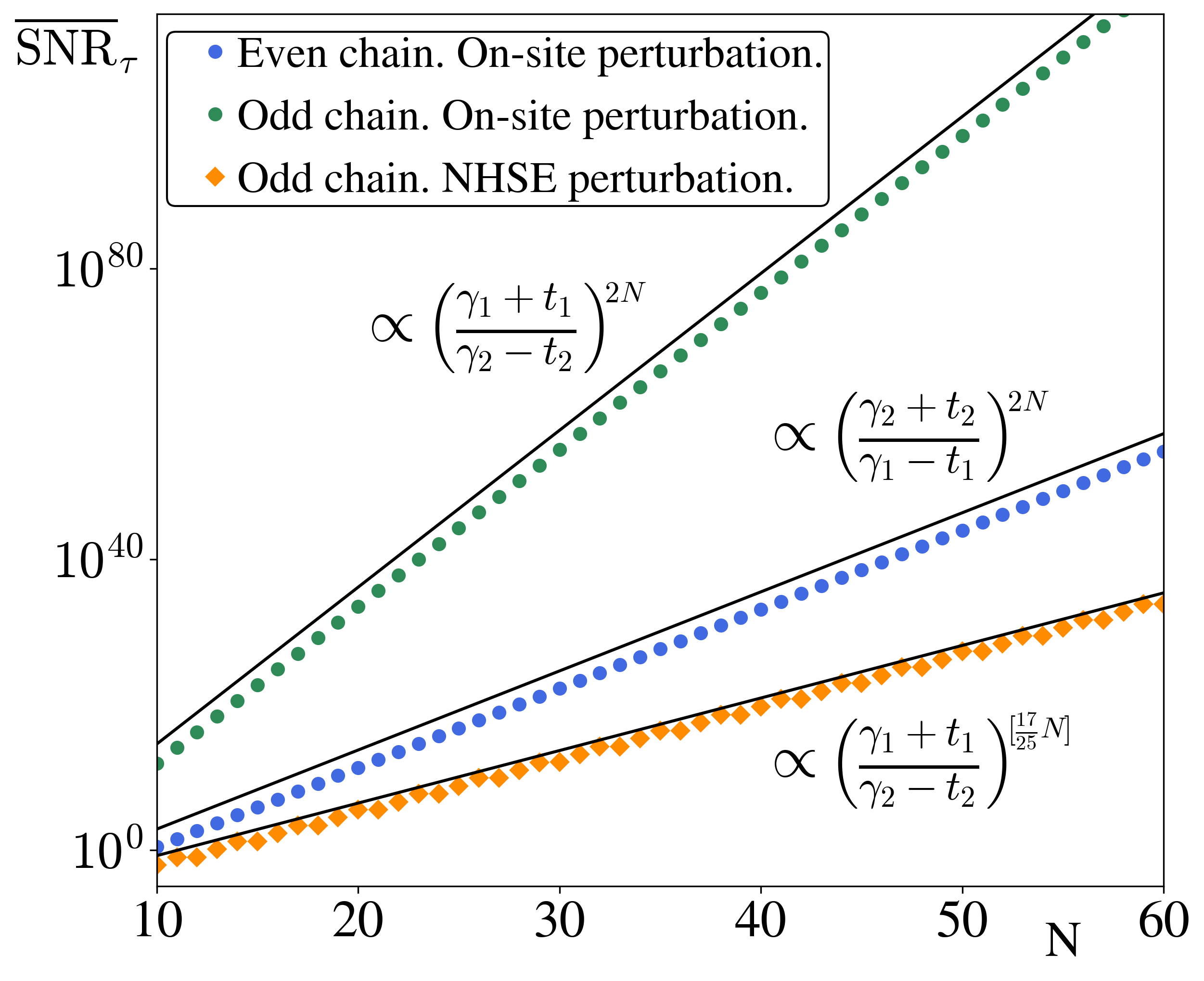}} 
    \caption{Scaling of the $\text{SNR}_\tau$ per number of photons with the system size $N$ (number of unit cells) for the on-site $\hat{V}_{N}$ and the NHSE perturbation $\hat{V}_{\text{NHSE}}$ at first order. Two main analysis are shown: (1) Comparison between odd and even chains with on-site perturbation, focusing on the regime where the odd chain outperforms the even chain. (2) Enhancement observed for the odd chain under NHSE perturbation, with no enhancement for the even chain (not shown).
    The parameters used for all plots are: $t_1= 0.5$, $t_2= 0.3$, $\gamma_1= 0.7$, $\gamma_2= 0.4$, $\theta= \frac{\pi}{4}$, $\kappa=0.05$, $\epsilon=10^{-6}$, $\varphi=0$, $\tau= 100$ and $n_{\text{th}}= 0$. For the on-site perturbation we set $m=1$ and for the NHSE perturbation we have $\phi = \frac{\pi}{2}$ and $m=[\alpha N]$, where $\alpha \approx \frac{9}{20}$ based on the previous parameters. Here, $[\alpha N]$ denotes the largest integer less than or equal to $\alpha N$.}
    \label{fig:SNRfirst}
\end{figure}

\section{SNR beyond the linear response regime}
Detecting infinitesimal parameters is insufficient for real-world applications. In this section, we no longer treat the perturbation to be infinitesimally small. Instead, we aim to understand how the results deviate from those in the linear regime. Our goal is to examine the SNR as the perturbation parameter $\epsilon$ changes from $\epsilon = 0$ to $\epsilon = \epsilon_0$, where $\epsilon_0$ is an arbitrary finite value. We will focus exclusively on the odd chain, which exhibits a single zero-energy edge mode under open boundary conditions. Terms beyond the zeroth order for both the noise and the number of photons will be considered, which also leads to their exponential amplification. Our figure of merit becomes
\begin{equation}
    \text{SNR}_\tau (N,\epsilon_0) \equiv \frac{|\langle \hat{\mathcal{M}}_\tau (N) \rangle_{\epsilon_0} - \langle \hat{\mathcal{M}}_\tau (N) \rangle_0 |^2}{\frac{\mathcal{N}_\tau (N,0)+\mathcal{N}_\tau (N,\epsilon_0)}{2}} .
    \label{eq:SNRAllDef}
\end{equation}
where we have averaged the noise at zeroth order $\mathcal{N}_\tau (N,0)$ and the noise at all orders $\mathcal{N}_\tau (N,\epsilon_0)$. Likewise, we take the total average number of photons to be 
\begin{equation}
    \bar{n}_{\text{tot}} \equiv \frac{\bar{n}_{\text{tot}}(0)+\bar{n}_{\text{tot}} (\epsilon_0)}{2} ,
\end{equation}
so that $\overline{\text{SNR}}_\tau (N,\epsilon_0) = \frac{\text{SNR}_\tau (N,\epsilon_0)}{\bar{n}_{\text{tot}}}$.

For the on-site perturbation $\hat{V}_{N} $, we fix $m=1$, $\varphi=0$ and $\theta = \frac{\pi}{2}$. As the system size increases, both the signal, $\mathcal{S}_\tau(N,\epsilon_0)$, and the number of photons, $\bar{n}_{\text{tot}} (N,\epsilon_0)$, grow exponentially with the system size $N$, with the signal scaling faster than the photon number. Initially, the output noise, $\mathcal{N}_\tau (N,\epsilon_0)$, remains at the vacuum noise level. However, once the system reaches an optimal size $N=N^*$ , the noise begins to grow at the same rate as the signal, causing the $\text{SNR}_\tau (N,\epsilon_0)$ to saturate. In the large $N$ limit, the $\text{SNR}_\tau (N,\epsilon_0)$ for our choice of parameters reads as
\begin{equation}
    \text{SNR}_\tau (N,\epsilon_0) \approx 8\tau\, \epsilon_0^{2}\,\kappa^2\,\abs{\beta}^2 \frac{1}{\mathcal{Q}} \left( \frac{\gamma_2\!+\!t_2}{\gamma_1\!-\!t_1}\right)^{\!\!4(N-1)} ,
\end{equation}
with
\begin{equation}
    \mathcal{Q} \approx 2\left(\frac{\kappa}{2}\right)^4 \!\left(\frac{\gamma_2^2-t_2^{2}}{\gamma_1^2-t_1^{2}}\right)^{\!\!4(N-1)}\!\! + \epsilon_0^2\,\kappa^2\left( \frac{\gamma_2\!+\!t_2}{\gamma_1\!-\!t_1}\right)^{\!\!4(N-1)} \!\!. \!\!\!
\end{equation}

\begin{figure}[H]
    \centering
    {\includegraphics[width=\linewidth]{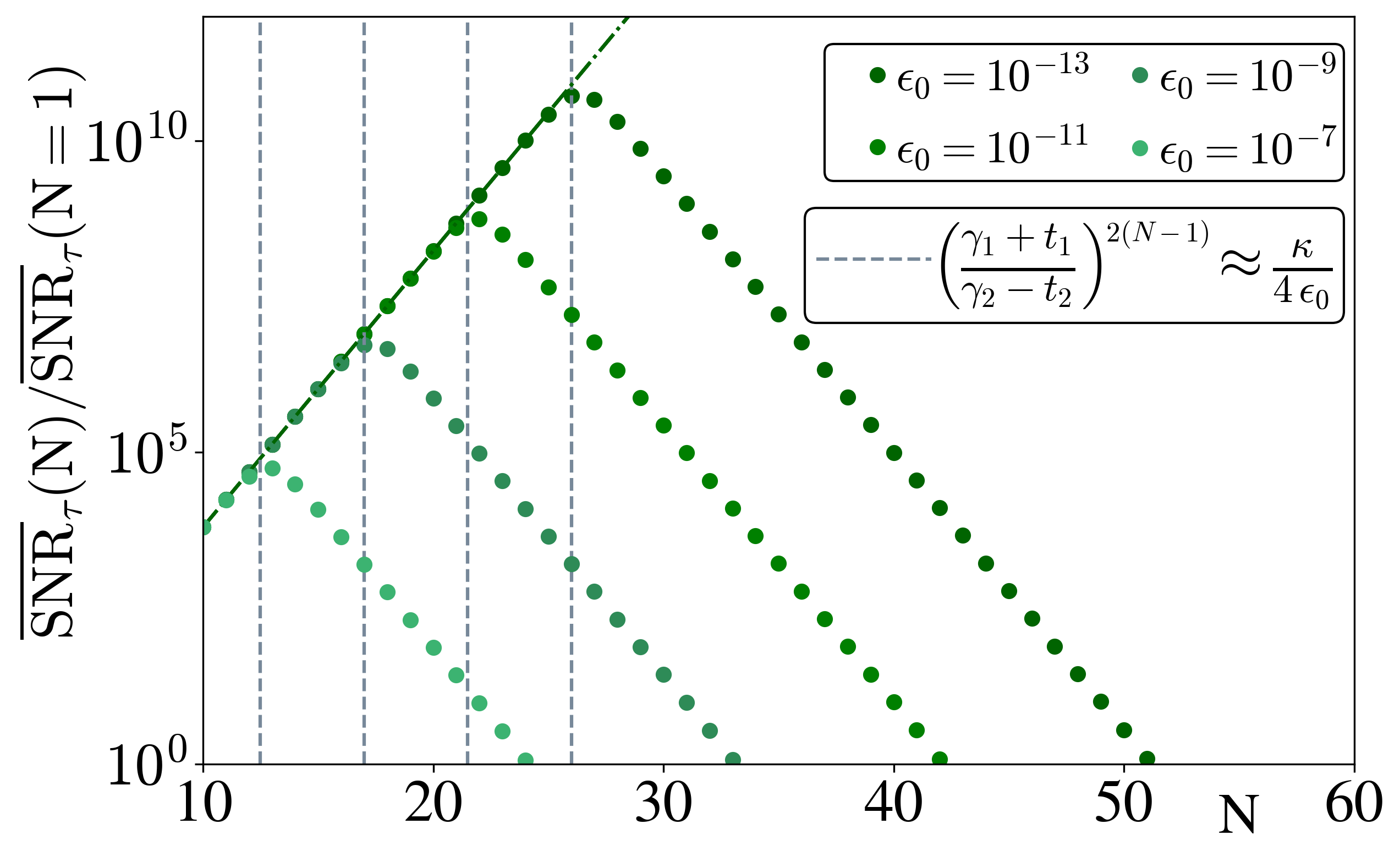}} 
    \caption{The plot shows the scaling of the signal-to-noise ratio $\overline{\text{SNR}}_\tau(N,\epsilon_0)/\overline{\text{SNR}}_\tau(N=1,\epsilon_0)$, as a function of $N$ (number of unit cells) for the on-site perturbation~$\hat{V}_{N}$, evaluated at various values of $\epsilon_0$. The initial exponential scaling aligns with the predictions from our linear response analysis (green dashed line). However, at higher orders, both noise and photon number are amplified, working together to prevent the signal-to-noise ratio from increasing indefinitely. The system size at which the linear regime breaks down is determined by the condition given in Eq.~\eqref{eq:conditionOnsite}. The parameters used are: $t_1 = t_2 = 1$, $\gamma_1 = 1.5$, $\gamma_2 = 2.5$, $m=1$, $\theta=~\frac{\pi}{2}$, $\kappa=~0.05$, $\varphi=~0$, $\tau=~100$ and $n_{\text{th}}=0$. Note that the plot is valid only for the odd chain.}
    \label{fig:SNRallOnsite}
\end{figure}

At first order, the first term in $\mathcal{Q}$ dominates. When $N=N^*$, the second term becomes dominant, causing the  $\text{SNR}_\tau (N,\epsilon_0)$ to saturate at $8\tau\abs{\beta}^2$. Notably, this upper bound of the $\text{SNR}_\tau (N,\epsilon_0)$ is independent of the perturbation strength $\epsilon_0$. Finally, when combined with the contribution from the amplification of the total number of photons (see the expression in Appendix~\ref{app:DynamicsAllOnsitePert}), this results in the behavior shown in FIG.~\ref{fig:SNRallOnsite}.

By examining the asymptotic behavior of the $\text{SNR}_\tau(N,\epsilon_0)$, we find that the linear regime breaks down when the following condition is satisfied
\begin{equation}
    \left( \frac{\gamma_1\!+\!t_1}{\gamma_2\!-\!t_2} \right)^{\!\!2(N^*-1)} \approx \frac{\kappa}{4 \,\epsilon_0} .
    \label{eq:conditionOnsite}
\end{equation}
Physically, this corresponds to the situation in which the output noise $\mathcal{N}_\tau (N,\epsilon_0)$ acquires twice the vacuum noise~\cite{McDonald2020}. We note that for fixed values of $t_1=t_2$ and $\gamma_1$, increasing $\gamma_2$ extends the linear regime to larger values of $N$ (though $\gamma_2$ cannot be increased indefinitely, as this would result in $N^*<1$). In that way, SSH-dynamics may allow the linear regime to apply to larger systems than those governed by Hatano-Nelson dynamics. 

\begin{figure}[H]
    \centering
    {\includegraphics[width=\linewidth]{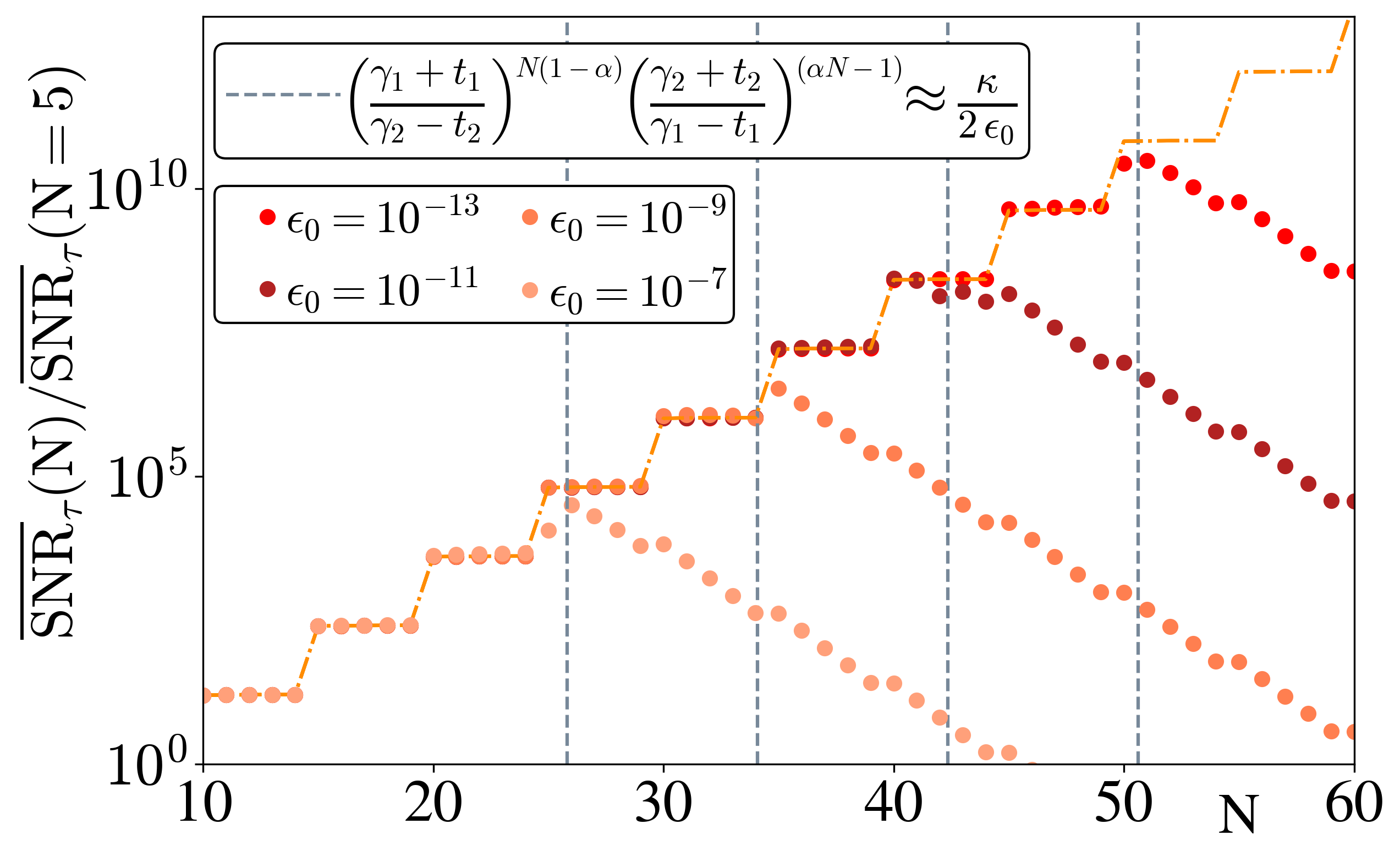}} 
    \caption{The plot shows the scaling of the signal-to-noise ratio $\overline{\text{SNR}}_\tau(N,\epsilon_0)/\overline{\text{SNR}}_\tau(N=5,\epsilon_0)$, as a function of $N$ (number of unit cells) for the NHSE perturbation~$\hat{V}_{\text{NHSE}}$, evaluated at various values of $\epsilon_0$. The initial exponential scaling is consistent with the predictions of our linear response analysis (orange dashed line). At higher orders, the saturation of the signal and the amplification of the photon number combines to limit the indefinite growth of the signal-to-noise ratio. The system size at which the linear regime breaks down is determined by the condition in Eq.~\eqref{eq:conditionNHSE}. The parameters used are: $t_1 = 0.6, t_2 = 0.4, \gamma_1 = 1.1, \gamma_2 = 1.6, \theta=~\frac{\pi}{4},\kappa=~0.05, \phi=\frac{\pi}{2},\varphi=~0,\tau=~100$, $n_{\text{th}}=0$ and $m=[\alpha N]$,  where $\alpha \approx \frac{1}{5}$ based on the previous parameters. Here, $[\alpha N]$ denotes the largest integer less than or equal to $\alpha N$. The plot is normalized to $N=5$ since this is the minimum size at which the NHSE perturbation shows exponential enhancement, given by  $N_{\text{min}} = \alpha^{-1}$. Note that the plot is valid only for the odd chain.}
    \label{fig:SNRallNHSE}
\end{figure}

We next study the NHSE perturbation $\hat{V}_{\text{NHSE}}$ and fix $\theta=\frac{\pi}{4}$, $\phi=\frac{\pi}{2}$ and $\varphi=0$. We let the position of the drive to be $m=\alpha N$, so that the $\text{SNR}_\tau (N,\epsilon_0)$ in this case is
\begin{equation}
    \!\!\!\text{SNR}_\tau (N,\epsilon_0)  \!=\!  \frac{8\tau \epsilon_0^{2}\abs{\beta}^2}{\mathcal{R}}\!\!\left( \!\frac{\gamma_2\!+\!t_2}{\gamma_1\!-\!t_1}\!\right)^{\!\!2(m-1)} \!\!\!\left( \!\frac{\gamma_1\!+\!t_1}{\gamma_2\!-\!t_2}\!\right)^{\!\!2(N-m)} \!\!\!\!\!, \!\!\!
\end{equation}
where
\begin{equation}
    \mathcal{R} = \left(\frac{\kappa}{2}\right)^2 \!+ \epsilon_0^2 \left( \!\frac{\gamma_2\!+\!t_2}{\gamma_1\!-\!t_1}\!\right)^{\!\!2(m-1)} \! \left( \!\frac{\gamma_1\!+\!t_1}{\gamma_2\!-\!t_2}\!\right)^{\!\!2(N-m)} .
\end{equation}
Analogously as we had for the on-site perturbation, we plot in FIG~\ref{fig:SNRallNHSE} the scaling of the $\text{SNR}_\tau (N,\epsilon_0)$ averaged with the contribution of the amplification of the total number photons. For the NHSE perturbation, the signal $\mathcal{S}_\tau(N,\epsilon_0)$ increases exponentially with the system size $N$ until it reaches an optimal size $N^*$, where it saturates. In contrast, the output noise $\mathcal{N}_\tau (N,\epsilon_0)$ exhibits a key distinction from the on-site perturbation: it remains constant at the vacuum level, even as $N$ increases. Note that the saturation behavior of the $\text{SNR}_\tau (N,\epsilon_0)$ for the NHSE perturbation is fundamentally different from that of the on-site perturbation. Here, saturation occurs because the signal itself reaches a limit, rather than due to the noise scaling to match the signal. At $N=~N^*$, the $\text{SNR}_\tau (N,\epsilon_0)$ shows a sharp peak before saturating at $8\tau\abs{\beta}^2$. This peak arises from a pole in the expressions for both the signal and noise (for details, see Appendix~\ref{app:DynamicsAllNHSEPert}), marking the breakdown of the linear regime. For the NHSE, the condition is given by
\begin{equation}
    \left( \frac{\gamma_1\!+\!t_1}{\gamma_2\!-\!t_2} \right)^{\!\!N^*(1-\alpha)} \left( \frac{\gamma_2\!+\!t_2}{\gamma_1\!-\!t_1}\right)^{\!\!\alpha N^*-1} \approx \frac{\kappa}{2 \,\epsilon_0} .
    \label{eq:conditionNHSE}
\end{equation}

We can again draw similar conclusions as as for the on-site perturbation. On one side, we see that the validity of the linear regime is controlled by the ratio $\frac{\kappa}{\epsilon_0}$. On the other side, for fixed values of $t_1 = t_2$ and $\gamma_1$, increasing $\gamma_2$ extends the linear regime to larger $N$, enabling SSH dynamics to apply to larger systems compared to those governed by Hatano-Nelson dynamics.

\section{Discussion}
In this study, we have explored a bosonic squeezed SSH model with parametric driving, focusing on its potential for sensing applications. The parametric driving is intrinsically connected to the system's NH dynamic properties. Our sensing targets have included two types of perturbations: a local on-site perturbation at the end of an open chain and a perturbation coupling the ends of the chain, related to the NHSE. To assess the sensor's performance, we have computed how the average signal-to-noise ratio, normalized by the total number of photons passing through the coupled resonators, depends on the system size. The normalization ensures that the observed enhancement with system size is not simply a result of the increased number of photons in the system but instead reflects the presence of a genuine enhancement mechanism. We have derived analytical expressions demonstrating an exponential dependence of the signal-to-noise ratio on system size, even when the total number of photons is held constant. 

For the detection of infinitesimal perturbations, we have compared our results for a chain with a broken last unit cell (odd number of sites) with those of a configuration featuring an intact unit cell at the chain's end (even number of sites), which exhibits different topological properties. Specifically, in the even-number-of-sites configuration, zero-energy edge modes appear only within specific parameter regimes. In contrast, the odd-number-of-sites setup always supports a zero-energy mode. This clarifies why a sensor with an odd number of sites can outperform its even-site counterpart in detecting on-site perturbations, particularly in regimes where edge modes are absent in the latter configuration. The difference becomes more pronounced when considering the NHSE perturbation: the signal-to-noise ratio from the sensor with an even number of sites is zero, while the configuration with an odd number of sites generates a finite response, provided the incident position of the driving field changes with the system size. We have then extended our analysis to regimes beyond linear response, examining the conditions under which the linear regime no longer applies. Finally, we have verified that the scheme discussed in this paper remains valid in the presence of local disorder, provided the disorder strength is smaller than the perturbation to be sensed, consistent with findings from recent work~\cite{FortinWang2024}. To ensure the robustness of our conclusions, all analytical findings have been thoroughly validated through numerical simulations. 

While previous studies have examined similar setups, there are significant distinctions between their approach and our results. A recent study has explored coupled resonators under SSH dynamics. However, two notable distinctions can be made. First, the earlier study examines a system with a complete last unit cell, where the dynamics are governed by an SSH model with alternating non-Hermitian and Hermitian couplings~\cite{ZhangChen}. These alternating couplings lead to dynamic instability in the system, as detailed in Appendix~\ref{app:stability}. We have addressed this issue by adopting the most general SSH model, where both couplings are non-Hermitian, within an appropriate parameter regime. Secondly, our focus has been placed on a chain with an odd number of sites, isolating the single zero-mode. This approach offers significant advantages when addressing the NHSE perturbation since, in a system with an even number of sites, enhancement through the NHSE can only be achieved by using protocols with multiple driving fields. In contrast, we have found here that for a chain with an odd number of sites, a single driving field suffices. Therefore, our results stand on the same footing as previous studies without the need for additional energy resources in the system, providing a direct generalization of their findings in the BKC~\cite{McDonald2020}. Exploring alternative protocols, such as using two driving fields, does not yield additional insights for an odd-number-of-sites chain and has therefore been omitted from this paper. Lastly, we stress that our sensing platform for the NHSE effect does not rely on fine-tuned parameter regimes.


Expanding on these results, we outline directions for future work. One potential approach could involve investigating deviations from the ideal system presented in this paper by examining the effects of gain and loss quantum noise, building on similar studies already conducted for the bosonic Kitaev chain~\cite{BaoQiNori2024}. Another promising direction is the generalization of these results beyond the input-output formalism, extending them to systems that require description through master equations in the presence of engineered reservoirs~\cite{MetelmannClerk2015,WangWangClerk2023,XieXu2024}.


\section*{Acknowledgements}
This work was supported by the Knut and Alice Wallenberg Foundation through the Wallenberg Center for Quantum Technology (WACQT), as well as the Wallenberg Academy Fellows (2018.0460) and Scholars (2023.0256) programs.

\clearpage


\clearpage
\pagebreak
\onecolumngrid
\widetext
\begin{center}
    \textbf{{APPENDICES}}\\[1em]
\end{center}

Here and in the main text, we adopt $\hbar=1$ and follow the conventions outlined in Ref.~\onlinecite{ClerkDevoret}. The appendices are structured as follows: Appendix~\ref{app:stability} demonstrates how the parameter regime with $\gamma_1<t_1$ and $\gamma_2<t_2$ results in dynamical instability. In Appendix~\ref{app:LinearResponse}, we solve the dynamics of the system for both for an infinitesimal on-site and NHSE perturbations, deriving and discussing the results presented in the main text. Appendix~\ref{app:FirstHpert} computes the matrix elements of $[\tilde{\mathds{H}}(\epsilon)]^{-1}$ using Dyson's series at first order. The matrix elements of $[\tilde{h}]^{-1}$ are provided in Appendix~\ref{app:tildeh}, and the relation between the matrices $h^X$, $h^P$ and $[\tilde{h}]$ is detailed in Appendix~\ref{app:hXPtildeh}. Finally, Appendix~\ref{app:BeyondLinearResponse} addresses the system's dynamics beyond linear response for both an on-site and NHSE perturbations, utilizing the matrix elements of $[\tilde{\mathds{H}}(\epsilon)]^{-1}$ at all orders listed in Appendix~\ref{app:AllHpert}.

\appendix
\section{Model with $\gamma_1<t_1$ and $\gamma_2<t_2$}
\label{app:stability}
We can study here the case the model with conditions $\gamma_1<t_1$ and $\gamma_2<t_2$. By applying the following squeezing transformation to Eq.~\eqref{eq:HamSystem}, 
\begin{equation}
\begin{alignedat}{2}
  \hat{x}_{n,A} &= e^{-\bar{r}(n-n_0)} \, e^{-\bar{s}(n-m_0)} \,\hat{\bar{x}}_{n,A} \,, \qquad &&\hat{x}_{n,B} = e^{-\bar{r}(n+1-n_0)}  \, e^{-\bar{s}(n-m_0)}  \,\hat{\bar{x}}_{n,B} \,, \\
  \hat{p}_{n,A} &= e^{\bar{r}(n-n_0)} \, e^{\bar{s}(n-m_0)} \,\hat{\bar{p}}_{n,A} \,, \qquad &&\hat{p}_{n,B} = e^{\bar{r}(n+1-n_0)} \, e^{\bar{s}(n-m_0)} \,\hat{\bar{p}}_{n,B} \,,
\end{alignedat}
\end{equation}
with $n_0, m_0$ being arbitrary constants and 
\begin{equation}
   \begin{alignedat}{2}
  e^{2\bar{r}} &\equiv \frac{t_1+\gamma_1}{t_1-\gamma_1} \,, \qquad \quad e^{2\bar{s}} &&\equiv \frac{t_2+\gamma_2}{t_2-\gamma_2} \,, \\
  \bar{t}_1 &\equiv \sqrt{t_1^2-\gamma_1^2}  \,, \qquad  \bar{t}_2 &&\equiv \sqrt{t_2^2-\gamma_2^2} \,,
  \end{alignedat}
\end{equation}
the Hamiltonian in quadrature basis is given by
\begin{equation}
     \begin{split}
    \hat{H}_S &= - \sum_{n=1}^{N-1}   \bar{t}_1 \, [ \hat{\bar{x}}_{n,A}\,\hat{\bar{p}}_{n,B} +  \hat{\bar{x}}_{n,B}\,\hat{\bar{p}}_{n,A}] -    \sum_{n=2}^{N} \bar{t}_2 \, [  \hat{\bar{x}}_{n,A}\,\hat{\bar{p}}_{n-1,B} +  \hat{\bar{x}}_{n-1,B}\,\hat{\bar{p}}_{n,A}] \\
    &= - \sum_{n=1}^{N-1}   i \, \bar{t}_1 \, [ \hat{\bar{a}}_{n,A}^\dagger\,\hat{\bar{a}}_{n,B}^\dagger -  \hat{\bar{a}}_{n,A}\,\hat{\bar{a}}_{n,N}] -    \sum_{n=2}^{N} i \,\bar{t}_2 \, [\hat{\bar{a}}_{n,A}^\dagger\,\hat{\bar{a}}_{n-1,B}^\dagger-\hat{\bar{a}}_{n,A}\,\hat{\bar{a}}_{n-1,B}] .
    \end{split}
\end{equation}
Thus, we can see that in the regime where $\gamma_1<t_1$ and $\gamma_2<t_2$, the model can be mapped by a similarity transformation to a model where only parametric driving is present with no hopping terms, resulting in unstable dynamics. In particular, this regime includes the case when $\gamma_2=0$~\cite{ZhangChen}.

\section{Dynamics in the Linear Response}
\label{app:LinearResponse}
We solve the dynamics of our model in the $\hat{x}$, $\hat{p}$ quadratures basis. The Heisenberg-Langevin equations~\eqref{EomInOut} are then given by
\begin{equation}
    \begin{split}
    \dot{\hat{x}}_{n,A} &= - (\gamma_1\!+\!t_1) \, \hat{x}_{n,B} + (\gamma_2\!-\!t_2) \, \hat{x}_{n-1,B} -i \,\epsilon \,[\hat{x}_{n,A},\hat{V}]  -\frac{\kappa}{2}\,\hat{x}_{m,A}\,\delta_{n m} - \sqrt{\kappa}\, (\beta_X \!+\! \hat{X}^{\text{in}})\,\delta_{n m} ,\\
    \dot{\hat{x}}_{n,B} &= +(\gamma_1\!-\!t_1) \, \hat{x}_{n,A} - (\gamma_2\!+\!t_2) \, \hat{x}_{n+1,A} ,\\
    \dot{\hat{p}}_{n,A} &= -(\gamma_1\!-\!t_1) \, \hat{p}_{n,B} + (\gamma_2\!+\!t_2) \, \hat{p}_{n-1,B} -i \,\epsilon \,[\hat{p}_{n,A},\hat{V}]  -\frac{\kappa}{2}\,\hat{p}_{m,A}\,\delta_{n m} - \sqrt{\kappa}\, (\beta_P \!+\! \hat{P}^{\text{in}})\,\delta_{n m} ,\\
    \dot{\hat{p}}_{n,B} &= +(\gamma_1\!+\!t_1) \, \hat{p}_{n,A} - (\gamma_2\!-\!t_2) \, \hat{p}_{n+1,A} ,
    \end{split}
    \label{EomInOutQuad}
\end{equation}
where we have defined $\beta \equiv \frac{1}{\sqrt{2}} (\beta_X \!+\! i \beta_P) = \abs{\beta}e^{i\theta}$ and $\hat{B}^{\text{in}}\equiv \frac{1}{\sqrt{2}} (\hat{X}^{\text{in}} \!+\! i \hat{P}^{\text{in}})$. The input noise quadratures $\hat{X}^{\text{in}}$ and $\hat{P}^{\text{in}}$ are completely determined by the first and second moments 
\begin{equation}
    \begin{split}
     \langle \hat{X}^{\text{in}}(t) \rangle = \langle \hat{P}^{\text{in}}(t) \rangle &= 0 ,\\
    \langle \hat{X}^{\text{in}}(t)\hat{X}^{\text{in} }(t') \rangle = \langle \hat{P}^{\text{in}}(t)\hat{P}^{\text{in}}(t') \rangle &= (\bar{n}_{\text{th}} \!+\! \frac{1}{2}) \,\delta(t-t') ,\\
    \langle \{\hat{X}^{\text{in}}(t), \hat{P}^{\text{in}}(t') \}\rangle &= 0 .
    \end{split}
\end{equation}
Defining $\mathbf{\hat{X}} = (\hat{x}_{1,A}, \hat{x}_{1,B}, \hat{x}_{2,A},\hat{x}_{2,B}, \dots, \hat{x}_{N,A})^T$ and $\mathbf{\hat{P}} = (\hat{p}_{1,A}, \hat{p}_{1,B}, \hat{p}_{2,A},\hat{p}_{2,B}, \dots, \hat{p}_{N,A})^T$ we can rewrite Eq.~\eqref{EomInOutQuad} in a matrix form 
\begin{equation}
    \begin{pmatrix}
     \mathbf{\dot{\hat{X}}} \\
     \mathbf{\dot{\hat{P}}} 
    \end{pmatrix}
    =\begin{pmatrix}
     h^X & 0 \\
      0  & h^P 
    \end{pmatrix}
    \begin{pmatrix}
     \mathbf{\hat{X}} \\
     \mathbf{\hat{P}} 
    \end{pmatrix}
    -i\,\epsilon
    \begin{pmatrix}
     [\mathbf{\hat{X}},\mathbf{\hat{V}}] \\
     [\mathbf{\hat{P}},\mathbf{\hat{V}}] 
    \end{pmatrix}
    -\boldsymbol{\beta} - \boldsymbol{\hat{\Omega}}^{\text{in}} ,
    \label{diffeqQuadMatrix}
\end{equation}
where the dynamical matrix is block-diagonal with~\cite{footnote}
\begin{equation}
    \begin{split}
    h^X = -\frac{\kappa}{2} \ket{m,A}\bra{m,A} &+ \sum_{n=1}^{N-1} \Bigl[ -(\gamma_1\!+\!t_1) \ket{n,A}\bra{n,B} + (\gamma_1\!-\!t_1) \ket{n,B}\bra{n,A}\Bigr) \Bigr] + \\
    &+\sum_{n=2}^{N} \Bigl[ (\gamma_2\!-\!t_2) \ket{n,A}\bra{n\!-\!1,B} - (\gamma_2\!-\!t_2)\ket{n\!-\!1,B}\bra{n,A} \Bigr] ,
    \end{split}
\end{equation}
and
\begin{equation}
    \begin{split}
    h^P = -\frac{\kappa}{2} \ket{m,A}\bra{m,A} &+ \sum_{n=1}^{N-1} \Bigl[ -(\gamma_1-t_1) \ket{n,A}\bra{n,B} + (\gamma_1\!+\!t_1) \ket{n,B}\bra{n,A} \Bigr] + \\
    &+\sum_{n=2}^{N} \Bigl[ (\gamma_2\!+\!t_2) \ket{n,A}\bra{n\!-\!1,B} - (\gamma_2\!-\!t_2)\ket{n\!-\!1,B}\bra{n,A} \Bigr] .
    \end{split}
\end{equation}
Note that for our choice of parameters, i.e., $\gamma_1>t_1, \gamma_2>t_2$ and $\kappa>0$, the eigenvalues of the dynamical matrix have negative real part (Routh-Hurwitz's stability criterion). The perturbation terms are
\begin{equation}
    \begin{pmatrix}
     [\mathbf{\hat{X}},\mathbf{\hat{V}}] \\
     [\mathbf{\hat{P}},\mathbf{\hat{V}}] 
    \end{pmatrix}
    = ([\hat{x}_{1,A}, \hat{V}],[\hat{x}_{1,B}, \hat{V}],\dots,[\hat{x}_{N,A}, \hat{V}],[\hat{p}_{1,A}, \hat{V}],[\hat{p}_{1,B}, \hat{V}],\dots [\hat{p}_{N,A}, \hat{V}]  )^T ,
\end{equation}
while the terms corresponding to the coherent driving and noise read
\begin{equation}
    \begin{split}
        \boldsymbol{\beta} &= \sqrt{2\kappa}\abs{\beta} \Bigl( \cos\theta \ket{m,A} + \sin\theta \ket{N\!+\!m,A} \Bigr) ,\\
        \boldsymbol{\hat{\Omega}}^{\text{in}} &= \sqrt{\kappa} \Bigl(\hat{X}^{\text{in}} \ket{m,A} +  \hat{P}^{\text{in}} \ket{N\!+\!m,A} \Bigr) .
    \end{split}
\end{equation}
Once the solution to the coupled differential equations in Eq.~\eqref{diffeqQuadMatrix} is obtained, we can evaluate the performance of our sensor. For that, it is useful to rearrange the expressions for the signal, noise and total number of photons in terms of the $\hat{x}$ and $\hat{p}$ quadratures. The signal $\mathcal{S}_\tau(N,\epsilon)$ 
 in Eq.~\eqref{SignalDef} for a protocol in which we drive the system with a coherent field and measure at the same $m$-mode in a sublattice $A$ reads as
\begin{equation}
    \mathcal{S}_\tau(N,\epsilon) = \kappa \tau \big| \cos\varphi \, (\langle\hat{x}_{m,A} \rangle_\epsilon - \langle\hat{x}_{m,A} \rangle_0) + \sin\varphi \, (\langle\hat{p}_{m,A} \rangle_\epsilon - \langle\hat{p}_{m,A} \rangle_0)\big|^2 ,
    \label{SignalProt1}
\end{equation}
where we have used Eq.~\eqref{Bout}, Eq.~\eqref{TempOutMode} and Eq.~\eqref{HomoCurrent}. Likewise, the noise $\mathcal{N}_\tau(N,\epsilon)$ defined in Eq.~\eqref{NoiseDef} reads as
\begin{equation}
     \mathcal{N}_\tau(N,\epsilon) = \langle \bigl(\cos\varphi \, \delta X + \sin\varphi \, \delta P\bigr)^2 \rangle_\epsilon ,
     \label{NoiseDefZero}
\end{equation}
with
\begin{equation}
    \begin{split}
          \delta X &\equiv  \frac{1}{\sqrt{\tau}} \int_0^\tau dt \,\Bigl[ \hat{X}^{\text{in}}(t) + \sqrt{\kappa} \, (\hat{x}_m(t) - \langle\hat{x}_m(t) \rangle_\epsilon)  \Bigr] ,\\
        \\
         \delta P &\equiv \frac{1}{\sqrt{\tau}} \int_0^\tau dt \,\Bigl[ 
         \hat{P}^{\text{in}}(t) + \sqrt{\kappa} \,(\hat{p}_m(t) - \langle\hat{p}_m(t) \rangle_\epsilon)  \Bigr] .
    \end{split}
\end{equation}
Finally, in the large-drive limit the total number of photons in Eq.~\eqref{PhotonsDef} is given via
\begin{equation}
    \bar{n}_{\text{tot}} = \frac{1}{2}\sum_{n=1}^N \Bigl(\langle \hat{x}_{n,A}\rangle^2_\epsilon + \langle \hat{p}_{n,A}\rangle^2_\epsilon + \langle \hat{x}_{n,B}\rangle^2_\epsilon + \langle \hat{p}_{n,B}\rangle^2_\epsilon \Bigr) .
    \label{PhotonsDefZero}
\end{equation}

\subsection{Solving the dynamics for the on-site perturbation.}
\label{app:DynamicsFirstOnsitePert}
The Heisenberg-Langevin equations~\eqref{EomInOutQuad} for the on-site perturbation $\epsilon \,\hat{V}_N = \epsilon \, \hat{a}_{N,A}^\dagger \hat{a}_{N,A} = \frac{\epsilon}{2}(\hat{x}_{N,A}\hat{x}_{N,A}+\hat{p}_{N,A}\hat{p}_{N,A})$ are given via
\begin{equation}
    \begin{split}
    \dot{\hat{x}}_{n,A} &= - (\gamma_1+t_1) \, \hat{x}_{n,B} + (\gamma_2-t_2) \, \hat{x}_{n-1,B} -\frac{\kappa}{2}\,\hat{x}_{m,A}\,\delta_{n m} +\epsilon \, \hat{p}_{N,A}\delta_{n N} - (\sqrt{2\kappa}\abs{\beta}\cos\theta + \sqrt{\kappa}\hat{X}^{\text{in}})\,\delta_{n m},\\
    \dot{\hat{x}}_{n,B} &= +( \gamma_1-t_1) \, \hat{x}_{n,A} - (\gamma_2+t_2) \, \hat{x}_{n+1,A} ,\\
    \dot{\hat{p}}_{n,A} &= -(\gamma_1-t_1) \, \hat{p}_{n,B} + (\gamma_2+t_2) \, \hat{p}_{n-1,B} -\frac{\kappa}{2}\,\hat{p}_{m,A}\,\delta_{n m} -\epsilon \, \hat{x}_{N,A}\delta_{n N} -  (\sqrt{2\kappa}\,\abs{\beta}\sin\theta + \sqrt{\kappa}\,\hat{P}^{\text{in}})\,\delta_{n m} ,\\
    \dot{\hat{p}}_{n,B} &= +(\gamma_1+t_1) \, \hat{p}_{n,A} - (\gamma_2-t_2) \, \hat{p}_{n+1,A} .
    \label{EomInOutQuadPert1}
    \end{split}
\end{equation}
One way to simplify the analysis of Eq.~\eqref{EomInOutQuadPert1} is by applying the following squeezing (Bogoliubov) transformation
\begin{equation}
\begin{alignedat}{2}
  \hat{x}_{n,A} &= e^{-r(n-n_0)} \, e^{-s(n-m_0)} \, \hat{\tilde{x}}_{n,A} \,, \qquad &&\hat{x}_{n,B} = e^{-r(n+1-n_0)}  \, e^{-s(n-m_0)}  \, \hat{\tilde{x}}_{n,B} \,,\\
  \hat{p}_{n,A} &= e^{r(n-n_0)} \, e^{s(n-m_0)} \,\hat{\tilde{p}}_{n,A} \,, \qquad &&\hat{p}_{n,B} = e^{r(n+1-n_0)} \, e^{s(n-m_0)} \,\hat{\tilde{p}}_{n,B} \,,
\end{alignedat}
\label{SqueezTrans}
\end{equation}
with $n_0,m_0$ being arbitrary constants and
\begin{equation}
   \begin{alignedat}{2}
   e^{2r} &\equiv \frac{\gamma_1+t_1}{\gamma_1-t_1} \,, \qquad \quad e^{2s} &&\equiv \frac{\gamma_2+t_2}{\gamma_2-t_2} \,, \\
  \tilde{t}_1 &\equiv \sqrt{\gamma_1^2-t_1^2}  \,, \qquad  \tilde{t}_2 &&\equiv \sqrt{\gamma_2^2-t_2^2} \,.
  \end{alignedat}
\end{equation}

An alternative, yet equivalent, approach would be to relate the matrix elements of the non-Hermitian $\tilde{h}^X$ and Hermitian $\tilde{h}^X$ matrices (provided in Appendix~\ref{app:hXPtildeh} for completeness), rather than performing the squeezing transformation. Remarkably, the above squeezing transformation transforms the Hamiltonian to
\begin{equation}
     \begin{split}
    \hat{H}_S &= \sum_{n=1}^{N-1} [ \tilde{t}_1 \, \hat{\tilde{x}}_{n,A}\,\hat{\tilde{p}}_{n,B} - \tilde{t}_1 \, \hat{\tilde{x}}_{n,B}\,\hat{\tilde{p}}_{n,A}] -  \sum_{n=2}^{N} [ \tilde{t}_2 \, \hat{\tilde{x}}_{n,A}\,\hat{\tilde{p}}_{n-1,B} - \tilde{t}_2 \,  \hat{\tilde{x}}_{n-1,B}\,\hat{\tilde{p}}_{n,A}] \\
    &=  \sum_{n=1}^{N-1} [ i \, \tilde{t}_1 \, \hat{\tilde{a}}_{n,A}\,\hat{\tilde{a}}_{n,B}^\dagger + \text{h.c.}] -  \sum_{n=2}^{N} [ i \, \tilde{t}_2 \, \hat{\tilde{a}}_{n,A}\,\hat{\tilde{a}}_{n-1,B}^\dagger + \text{h.c.}]  ,
    \end{split}
\end{equation}
i.e., in the stable regime where $\gamma_1>t_1$ and $\gamma_2>t_2$, we find that the Hamiltonian is equivalent to a tight-binding model with hopping terms $i \, \tilde{t}_1, i \, \tilde{t}_2$ and no parametric driving. Note also that the Hamiltonian is independent of the constants $n_0,m_0$. The dynamical matrix and the equations of motions in Eq.~\eqref{EomInOutQuadPert1} become
\begin{equation}
    \begin{split}
    \dot{\hat{\tilde{x}}}_{n,A} &= - \tilde{t}_1  \, \hat{\tilde{x}}_{n,B} + \tilde{t}_2 \, \hat{\tilde{x}}_{n-1,B} -\frac{\kappa}{2}\,\hat{\tilde{x}}_{m,A}\,\delta_{n m} +\epsilon \, \hat{\tilde{p}}_{N,A}\, e^{2r(N-n_0)}\,e^{2s(N-m_0)}\,\delta_{n N} ,\\
    &\qquad - (\sqrt{2\kappa}\abs{\beta}\cos\theta + \sqrt{\kappa}\hat{X}^{\text{in}})\,e^{r(m-n_0)}\,e^{s(m-m_0)}\,\delta_{n m} ,\\
    \dot{\hat{\tilde{x}}}_{n,B} &= +\tilde{t}_1  \, \hat{\tilde{x}}_{n,A} - \tilde{t}_2 \, \hat{\tilde{x}}_{n+1,A} ,\\
    \dot{\hat{\tilde{p}}}_{n,A} &= -\tilde{t}_1 \, \hat{\tilde{p}}_{n,B} + \tilde{t}_2 \, \hat{\tilde{p}}_{n-1,B} -\frac{\kappa}{2}\,\hat{\tilde{p}}_{m,A}\,\delta_{n m} -\epsilon \, \hat{\tilde{x}}_{N,A}\,e^{-2r(N-n_0)}\, e^{-2s(N-m_0)}\,\delta_{n N} ,\\
    &\qquad - (\sqrt{2\kappa}\,\abs{\beta}\sin\theta + \sqrt{\kappa}\,\hat{P}^{\text{in}})\,e^{-r(m-n_0)}\,e^{-s(m-m_0)}\,\delta_{n m} ,\\
    \dot{\hat{\tilde{p}}}_{n,B} &= +\tilde{t}_1 \, \hat{\tilde{p}}_{n,A} - \tilde{t}_2 \, \hat{\tilde{p}}_{n+1,A} .
    \label{EomInOutQuadPert1Simp}
    \end{split}
\end{equation}
We can study the dynamics in the long-time limit by Fourier transforming Eq.~\eqref{EomInOutQuadPert1Simp}. Defining the susceptibility matrix $\chi[\omega,\epsilon] = (-i \omega \mathds{1}-\tilde{\mathds{H}}_{N}(\epsilon))^{-1}$ where $\tilde{\mathds{H}}_{N}(\epsilon) = \tilde{\mathds{H}}(\kappa)+\tilde{\mathds{H}}_{\text{pert},N}(\epsilon)$ with
\begin{equation}
    \begin{split}
    \tilde{\mathds{H}}(\kappa) &= -\frac{\kappa}{2} \ket{m,A}\bra{m,A} -\frac{\kappa}{2} \ket{2N\!+\!2m\!-\!1,A}\bra{2N\!+\!2m\!-\!1,A} + \\
    &+ \sum_{n=1}^{N-1} \Bigl[ \tilde{t}_1 \Bigl(-\ket{n,A}\bra{n,B} + \ket{n,B}\bra{n,A}\Bigr) \Bigr] +\sum_{n=2}^{N} \Bigl[ \tilde{t}_2 \Bigl(\ket{n,A}\bra{n\!-\!1,B} - \ket{n\!-\!1,B}\bra{n,A}\Bigr) \Bigr]+ \\
    &+ \sum_{n=N+1}^{2N-1} \Bigl[ \tilde{t}_1 \Bigl(-\ket{n,A}\bra{n,B} + \ket{n,B}\bra{n,A}\Bigr) \Bigr] +\sum_{n=N+2}^{2N} \Bigl[ \tilde{t}_2 \Bigl(\ket{n,A}\bra{n\!-\!1,B} - \ket{n\!-\!1,B}\bra{n,A}\Bigr) \Bigr] ,
    \end{split}
    \label{Hk}
\end{equation}
which is a block-diagonal matrix that consists of two $(2N\!-\!1)$-dimensional blocks ($\tilde{h}$ in Appendix~\ref{app:tildeh}) and
\begin{equation}
    \tilde{\mathds{H}}_{\text{pert},N}(\epsilon) = \epsilon \Bigl(e^{2r(N-n_0)}e^{2s(N-m_0)}-  \ket{N,A}\bra{2N,A} - e^{-2r(N-n_0)} e^{-2s(N-m_0)}\ket{2N,A}\bra{N,A} \Bigr) \,.
\end{equation}
Here we restrict to the zero-frequency susceptibility matrix $\chi[0,\epsilon] = (-\tilde{\mathds{H}}_{N}[\epsilon])^{-1}$. The dynamics are given via
\begin{equation}
    \begin{split}
    \hat{\tilde{x}}_{n,A} =& \, [\tilde{\mathds{H}}_{N}(\epsilon)]^{-1}_{2n-1,2m-1} \, (\sqrt{2\kappa}\abs{\beta}\cos\theta + \sqrt{\kappa}\hat{X}^{\text{in}})\,e^{r(m-n_0)} \, e^{s(m-m_0)} \,+ \\ 
    &+[\tilde{\mathds{H}}_{N}(\epsilon)]^{-1}_{2n-1,2N+2m-2} \,(\sqrt{2\kappa}\,\abs{\beta}\sin\theta + \sqrt{\kappa}\,\hat{P}^{\text{in}})\,e^{-r(m-n_0)}\,e^{-s(m-m_0)} ,\\
    \hat{\tilde{x}}_{n,B} =& \, [\tilde{\mathds{H}}_{N}(\epsilon)]^{-1}_{2n,2m-1} \, (\sqrt{2\kappa}\abs{\beta}\cos\theta + \sqrt{\kappa}\hat{X}^{\text{in}})\,e^{r(m-n_0)} \,e^{s(m-m_0)} \, + \\
    &+[\tilde{\mathds{H}}_{N}(\epsilon)]^{-1}_{2n,2N+2m-2} \,(\sqrt{2\kappa}\,\abs{\beta}\sin\theta + \sqrt{\kappa}\,\hat{P}^{\text{in}})\,e^{-r(m-n_0)}\,e^{-s(m-m_0)} ,\\    
    \hat{\tilde{p}}_{n,A} =&  \, [\tilde{\mathds{H}}_{N}(\epsilon)]^{-1}_{2N+2n-2,2m-1} \, (\sqrt{2\kappa}\abs{\beta}\cos\theta + \sqrt{\kappa}\hat{X}^{\text{in}})\,e^{r(m-n_0)} \, e^{s(m-m_0)} \,+\\
    &+[\tilde{\mathds{H}}_{N}(\epsilon)]^{-1}_{2N+2n-2,2N+2m-2} \,(\sqrt{2\kappa}\,\abs{\beta}\sin\theta + \sqrt{\kappa}\,\hat{P}^{\text{in}})\,e^{-r(m-n_0)}\,e^{-s(m-m_0)} ,\\
    \hat{\tilde{p}}_{n,B} =& \, [\tilde{\mathds{H}}_{N}(\epsilon)]^{-1}_{2N+2n-1,2m-1} \, (\sqrt{2\kappa}\abs{\beta}\cos\theta + \sqrt{\kappa}\hat{X}^{\text{in}})\,e^{r(m-n_0)} \, e^{s(m-m_0)} \, +\\
    &+ [\tilde{\mathds{H}}_{N}(\epsilon)]^{-1}_{2N+2n-1,2N+2m-2} \,(\sqrt{2\kappa}\,\abs{\beta}\sin\theta + \sqrt{\kappa}\,\hat{P}^{\text{in}})\,e^{-r(m-n_0)}\,e^{-s(m-m_0)} ,
    \end{split}
    \label{eq:QuadPert1Simp}
\end{equation}
where $1\leq n \leq N$ with $\hat{\tilde{x}}_{N,B} =  \hat{\tilde{p}}_{N,B} \equiv 0$.

For an \textit{infinitesimal} $\epsilon$, we can compute the matrix elements of $[\tilde{\mathds{H}}_{N}(\epsilon)]^{-1}$ at first order using Dyson's equation in terms of the matrix elements of $[\tilde{h}]^{-1}$ of the dynamical matrix (see Appendix~\ref{app:FirstHpert}). Then, applying the inverse of the squeezing transformation in Eq.~\eqref{SqueezTrans} is straightforward to get $\hat{x}_{n,A}, \hat{x}_{n,B}, \hat{p}_{n,A}$ and $\hat{p}_{n,B}$. The relevant expectation values for the signal in Eq.~\eqref{SignalProt1} can be expressed as
\begin{equation}
    \langle\hat{x}_{m,A} \rangle_\epsilon - \langle\hat{x}_{m,A} \rangle_0 = -\epsilon \left( \frac{2}{\kappa}\right)^{\!\!2} \left( \frac{\gamma_1\!+\!t_1}{\gamma_2\!-\!t_2}\right) ^{\!\!2(N-m)} \, \sqrt{2\kappa} \, \abs{\beta} \, \sin\theta ,
\end{equation}
and
\begin{equation}
    \langle\hat{p}_{m,A} \rangle_\epsilon - \langle\hat{p}_{m,A} \rangle_0 = \epsilon \left( \frac{2}{\kappa}\right)^{\!\!2} \left( \frac{\gamma_1\!-\!t_1}{\gamma_2\!+\!t_2}\right) ^{\!\!2(N-m)} \, \sqrt{2\kappa} \, \abs{\beta} \, \cos\theta ,
\end{equation}
so that the signal is given by
\begin{equation}
    \mathcal{S}_\tau(N,\epsilon) = 32 \tau \frac{\epsilon^2}{\kappa^2}\abs{\beta}^2 \Big[ -\cos\varphi \, \sin\theta \left( \frac{\gamma_1\!+\!t_1}{\gamma_2\!-\!t_2}\right)^{\!\!2(N-m)} + \sin\varphi \, \cos\theta \, \left( \frac{\gamma_1\!-\!t_1}{\gamma_2\!+\!t_2}\right) ^{\!\!2(N-m)} \Big]^2 .
    \label{SignalFirstOnsiteFP}
\end{equation}
Since we are keeping terms up to first order in the SNR we are only concerned with the noise in Eq.~\eqref{NoiseDefZero} at zeroth order, which is given by
\begin{equation}
    \begin{split}
    \mathcal{N}_\tau(N,0) &= \langle \bigl(\cos\varphi \, \frac{1}{\sqrt{\tau}}\int_0^\tau dt \, (1+\kappa\, [h^X]^{-1}_{2m-1,2m-1}) \hat{X}^{\text{in}} + \sin\varphi   \, \frac{1}{\sqrt{\tau}}\int_0^\tau dt \, (1+\kappa\, [h^P]^{-1}_{2m-1,2m-1}) \hat{P}^{\text{in}}   \bigr)^2 \rangle_0 \\
    &= \Bigl( \cos^2\varphi \, (1+\kappa\, [h^X]^{-1}_{2m-1,2m-1})^2 + \sin^2\varphi \, (1+\kappa\, [h^P]^{-1}_{2m-1,2m-1})^2 \Bigr) \, ( \bar{n}_{\text{th}} + \frac{1}{2}) = \bar{n}_{\text{th}}+ \frac{1}{2} ,
    \label{NoiseZerothFP}
    \end{split}
\end{equation}
where we have used that $[h^X]^{-1}_{2m-1,2m-1} = [h^P]^{-1}_{2m-1,2m-1}  = \frac{2}{\kappa}$ and $\bar{n}_{\text{th}}$ is the thermal occupation number of the vacuum noise. Likewise, we compute the total number of photons in Eq.~\eqref{PhotonsDefZero} at zeroth order as
\begin{equation}
    \bar{n}_{\text{tot}}(0) = \frac{1}{2}\sum_{n=1}^N \Bigl(\langle \hat{x}_{n,A}\rangle^2_0 + \langle \hat{p}_{n,A}\rangle^2_0 \Bigr)=\kappa \abs{\beta}^2 \sum_{n=1}^N \Bigl[\cos^2\theta \, ([h^X]^{-1}_{2n-1,2m-1})^2  + \sin^2\theta \, ([h^P]^{-1}_{2n-1,2m-1})^2   \Bigr] .
\end{equation}
At zeroth order, we note that only the sites within sublattice $A$ contribute, unlike for the case of a chain with an \textit{even} number of sites, where only the $B$-sites are populated at zeroth order. The number of photons in terms of the system parameters is given by
\begin{equation}
    \bar{n}_{\text{tot}}(0) = \frac{4}{\kappa} \abs{\beta}^2 \left\{ \cos^2\theta \, \Bigl[ \frac{\gamma_2\!+\!t_2}{\gamma_1\!-\!t_1}\Bigr]^{2(m-1)} \,\left(\frac{1-\left(\frac{\gamma_1-t_1}{\gamma_2+t_2} \right)^{\!2N}}{1-\left(\frac{\gamma_1-t_1}{\gamma_2+t_2} \right)^{\!2}}\right) + \sin^2\theta \, \Bigl[ \frac{\gamma_2\!-\!t_2}{\gamma_1\!+\!t_1}\Bigr]^{2(m-1)} \,\left(\frac{1-\left(\frac{\gamma_1+t_1}{\gamma_2-t_2} \right)^{\!2N}}{1-\left(\frac{\gamma_1+t_1}{\gamma_2-t_2} \right)^{\!2}}  \right) \right\} .
    \label{PhotonsZerothFP}
\end{equation}

In \textit{regime I}, where $\abs{\gamma_2\!+\!t_2}>\abs{\gamma_1\!-\!t_1}$ and $\abs{\gamma_2\!-\!t_2}>\abs{\gamma_1\!+\!t_1}$, and in the large N-limit we find that
\begin{equation}
    \overline{\text{SNR}}_\tau (N) \, \approx  \frac{8\tau\epsilon^2}{\kappa\,(\frac{1}{2}\!+\!\bar{n}_{\text{th}})}  \, \frac{\Big[ -\cos\varphi \, \sin\theta \left( \frac{\gamma_1+t_1}{\gamma_2-t_2}\right)^{\!2(N-m)} + \sin\varphi \, \cos\theta \, \left( \frac{\gamma_1-t_1}{\gamma_2+t_2}\right) ^{\!2(N-m)} \Big]^2}{\cos^2\theta \, \left( \frac{\gamma_2+t_2}{\gamma_1-t_1}\right)^{\!2(m-1)} + \sin^2\theta \left( \frac{\gamma_2-t_2}{\gamma_1+t_1}\right)^{\!2(m-1)}}  .
\end{equation}
There is exponential attenuation with $\overline{\text{SNR}}_\tau (N) \!\propto\! \left( \frac{\gamma_1+t_1}{t_2-\gamma_2}\right)^{\!4N}$ for $\varphi\in[0,\frac{\pi}{2})$ and $\theta \neq \{0,\pi\}$. In the fine-tuned regime where $\varphi=\frac{\pi}{2}$ and $\theta \neq \frac{\pi}{2}$ we also find exponential attenuation with $\overline{\text{SNR}}_\tau (N) \!\propto\! \left( \frac{\gamma_1-t_1}{\gamma_2+t_2}\right)^{\!4N}$.

In \textit{regime II}, where $\abs{\gamma_2\!+\!t_2}>\abs{\gamma_1\!-\!t_1}$ and $\abs{\gamma_2\!-\!t_2}<\abs{\gamma_1\!+\!t_1}$, and in the large N-limit we find that
\begin{equation}
    \overline{\text{SNR}}_\tau (N) \, \approx  \frac{8\tau\epsilon^2\, \cos^2\varphi}{\kappa\,(\frac{1}{2}\!+\!\bar{n}_{\text{th}})}  \, \frac{\sin^2\theta \, \left(\frac{\gamma_1+t_1}{\gamma_2-t_2}\right)^{\!4(N-m)}}{\cos^2\theta \, \left( \frac{\gamma_2+t_2}{\gamma_1-t_1}\right)^{\!2(m-1)} + \sin^2\theta \left( \frac{\gamma_1+t_1}{\gamma_2-t_2}\right)^{\!2(N-m)}}  .
\end{equation}
We find exponential amplification with $\overline{\text{SNR}}_\tau (N) \!\propto\! \left( \frac{\gamma_1+t_1}{\gamma_2-t_2}\right)^{\!2N}$ for $\varphi\in[0,\frac{\pi}{2})$ and $\theta \neq \{0,\pi\}$. However, exponential attenuation also exists for fine-tuned points in parameter space with $\overline{\text{SNR}}_\tau (N) \!\propto\! \left( \frac{\gamma_1-t_1}{\gamma_2+t_2}\right)^{\!4N}$for $\varphi\in(0,\frac{\pi}{2}]$ and ${\theta=\{0,\pi\}}$ and $\overline{\text{SNR}}_\tau (N) \!\propto\! \frac{\left( \frac{\gamma_1-t_1}{\gamma_2+t_2}\right)^{\!4N}}{\left( \frac{\gamma_1+t_1}{\gamma_2-t_2}\right)^{\!2N}}$ for $\varphi=\frac{\pi}{2}$ and $\theta \neq \{0,\frac{\pi}{2},\pi\}$.

In \textit{regime III}, where $\abs{\gamma_2\!+\!t_2}<\abs{\gamma_1\!-\!t_1}$ and $\abs{\gamma_2\!-\!t_2}<\abs{\gamma_1\!+\!t_1}$, and in the large N-limit we find that
\begin{equation}
    \overline{\text{SNR}}_\tau (N) \approx  \frac{8\tau\epsilon^2}{\kappa\,(\frac{1}{2}\!+\!\bar{n}_{\text{th}})}   \, \frac{\Big[ -\cos\varphi \, \sin\theta \left( \frac{\gamma_1+t_1}{\gamma_2-t_2}\right)^{\!2(N-m)} + \sin\varphi \, \cos\theta \, \left( \frac{\gamma_1-t_1}{\gamma_2+t_2}\right) ^{\!2(N-m)} \Big]^2 }{\cos^2\theta \, \left( \frac{\gamma_1-t_1}{\gamma_2+t_2}\right)^{\!2(N-m)} + \sin^2\theta \left( \frac{\gamma_1+t_1}{\gamma_2-t_2}\right)^{\!2(N-m)}} .
\end{equation}
In this regime, we mostly find exponential amplification scaling as $\overline{\text{SNR}}_\tau (N) \!\propto\! \left( \frac{\gamma_1+t_1}{\gamma_2-t_2}\right)^{\!2N}$ for $\varphi\in [0,\frac{\pi}{2})$ and $\theta \neq\{0,\pi\}$. Fine-tuned exponential amplification can also be found with $\overline{\text{SNR}}_\tau (N) \!\propto\! \left( \frac{\gamma_1-t_1}{\gamma_2+t_2}\right)^{\!2N}$ for $\varphi=\frac{\pi}{2}$ and $\theta =\{0,\pi\}$, $\theta\neq \frac{\pi}{2}$. Lastly, there is fine-tuned exponential attenuation with
 $\overline{\text{SNR}}_\tau (N) \!\propto\! \frac{\left( \frac{\gamma_1-t_1}{\gamma_2+t_2}\right)^{\!4N}}{\left( \frac{\gamma_1+t_1}{\gamma_2-t_2}\right)^{\!2N}}$ for $\varphi=\frac{\pi}{2}$ and $\theta \neq \{0,\frac{\pi}{2},\pi\}$. 

Finally, for comparison, we provide the results for the even chain case. The signal at first order is given via~\cite{ZhangChen}
\begin{equation}
    \mathcal{S}^{\text{even}}_\tau(N,\epsilon) = 2\tau \kappa^2 \epsilon^2\abs{\beta}^2 \Big[ \cos\varphi \, \sin\theta  \frac{1}{(\gamma_1\!-\!t_1)^2}\left( \frac{\gamma_2\!+\!t_2}{\gamma_1\!-\!t_1}\right)^{\!\!2(N-m)} - \sin\varphi \, \cos\theta \, \frac{1}{(\gamma_1\!+\!t_1)^2}\left( \frac{\gamma_2\!-\!t_2}{\gamma_1\!+\!t_1}\right)^{\!\!2(N-m)}\Big]^2 .
\end{equation}
The noise and the number of photons at zeroth order is, respectively, $ \mathcal{N}_\tau(N,0) = \bar{n}_{\text{th}}+\frac{1}{2}$ and~\cite{ZhangChen}
\begin{equation}
     \bar{n}_{\text{tot}}^{\text{even}}(0) = \kappa \abs{\beta}^2 \Bigl\{ \cos^2\theta \,  \frac{1}{(\gamma_1\!+\!t_1)^2} \frac{1-\left(\frac{\gamma_2-t_2}{\gamma_1+t_1} \right)^{\!2(N-m+1)}}{1-\left(\frac{\gamma_2-t_2}{\gamma_1+t_1} \right)^{\!2}} + \sin^2\theta \, \frac{1}{(\gamma_1\!-\!t_1)^2} \frac{1-\left(\frac{\gamma_2+t_2}{\gamma_1-t_1} \right)^{\!2(N-m+1)}}{1-\left(\frac{\gamma_2+t_2}{\gamma_1-t_1} \right)^{\!2}}  \Bigr\} .
\end{equation}
For the parameters $m=1, \varphi=0$ and $\theta=\frac{\pi}{2}$, the even chain in \textit{Regime I} and \textit{Regime II} has signal and number of photons given via
\begin{equation}
    \begin{split}
    \mathcal{S}^{\text{even}}_\tau(N,\epsilon) &= 2\tau \kappa^2 \epsilon^2\abs{\beta}^2   \frac{1}{(\gamma_1\!-\!t_1)^4}\left( \frac{\gamma_2+t_2}{\gamma_1-t_1}\right)^{\!\!4(N-1)} \,, \\ 
    \bar{n}_{\text{tot}}^{\text{even}}(0) &\approx \kappa \abs{\beta}^2 \frac{1}{(\gamma_1-t_1)^2} \left(\frac{\gamma_2+t_2}{\gamma_1-t_1} \right)^{\!\!2(N-1)} ,
    \end{split}
\end{equation}
so that we have exponential enhancement with
\begin{equation}
    \text{SNR}_\tau^{\text{even}}(N,\epsilon) = 4 \, \frac{\kappa\tau\,\bar{n}^{\text{even}}_{\text{tot}}}{(2\, n_{\text{th}}\!+\!1)} \frac{\epsilon^2}{(\gamma_1\!-\!t_1)^2} \left( \frac{\gamma_2\!+\!t_2}{\gamma_1\!-\!t_1} \right)^{\!\!2(N-1)}  .
\end{equation}

\subsection{Solving the dynamics for the NHSE perturbation.}
\label{app:DynamicsFirstNHSEPert}

The Heisenberg-Langevin equations~\eqref{EomInOutQuad} for the NHSE perturbation $\epsilon \hat{V}_{\text{NHSE}} = \epsilon (e^{i\phi}\,\hat{a}_{1,A} \hat{a}_{N,A}^\dagger + e^{-i\phi}\,\hat{a}_{1,A}^\dagger \hat{a}_{N,A})= \epsilon\,\cos\phi \,(\hat{x}_{1,A}\hat{x}_{N,A} + \hat{p}_{1,A}\hat{p}_{N,A} ) -\epsilon\, \sin\phi \, (\hat{p}_{1,A}\hat{x}_{N,A} - \hat{x}_{1,A}\hat{p}_{N,A})$ are given via
\begin{equation}
    \begin{split}
    \dot{\hat{x}}_{n,A} &= - (\gamma_1\!+\!t_1) \, \hat{x}_{n,B} + (\gamma_2\!-\!t_2) \, \hat{x}_{n-1,B} -\frac{\kappa}{2}\,\hat{x}_{m,A}\,\delta_{n m} - (\sqrt{2\kappa}\abs{\beta}\cos\theta + \sqrt{\kappa}\hat{X}^{\text{in}})\,\delta_{n m}  \\
    &+\epsilon \cos\phi \, (\hat{p}_{N,A} \, \delta_{n 1}+\hat{p}_{1,A} \,\delta_{n N}) -\epsilon \sin\phi \,(\hat{x}_{N,A}\, \delta_{n 1}-\hat{x}_{1,A} \, \delta_{n N}) ,\\
        \dot{\hat{x}}_{n,B} &= +(\gamma_1\!-\!t_1) \, \hat{x}_{n,A} - (\gamma_2\!+\!t_2) \, \hat{x}_{n+1,A} ,\\
    \dot{\hat{p}}_{n,A} &= -(\gamma_1\!-\!t_1) \, \hat{p}_{n,B} + (\gamma_2\!+\!t_2) \, \hat{p}_{n-1,B} -\frac{\kappa}{2}\,\hat{p}_{m,A}\,\delta_{n m} -  (\sqrt{2\kappa}\,\abs{\beta}\sin\theta + \sqrt{\kappa}\,\hat{P}^{\text{in}})\,\delta_{n m} \\
    &-\epsilon \cos\phi \,(\hat{x}_{N,A} \,\delta_{n 1}+\hat{x}_{1,A} \,\delta_{n N}) +\epsilon \sin\phi \,(\hat{p}_{1,A} \,\delta_{n N}-\hat{p}_{N,A} \,\delta_{n 1}) ,\\
    \dot{\hat{p}}_{n,B} &= +(\gamma_1\!+\!t_1) \, \hat{p}_{n,A} - (\gamma_2\!-\!t_2) \, \hat{p}_{n+1,A} .
    \label{EomInOutQuadPert2}
    \end{split}
\end{equation}
Applying the squeezing transformation in Eq.~\eqref{SqueezTrans} to Eq.~\eqref{EomInOutQuadPert2}, leads to the equivalent dynamics described via the equations
\begin{equation}
    \begin{split}
    \dot{\hat{\tilde{x}}}_{n,A} &= - \tilde{t}_1  \, \hat{\tilde{x}}_{n,B} + \tilde{t}_2 \, \hat{\tilde{x}}_{n-1,B} -\frac{\kappa}{2}\,\hat{\tilde{x}}_{m,A}\,\delta_{n m} - (\sqrt{2\kappa}\abs{\beta}\cos\theta + \sqrt{\kappa}\hat{X}^{\text{in}})\, e^{r(m-n_0)}\,e^{s(m-m_0)}\,\delta_{n m}  \\
    &\quad +\epsilon \cos\phi \, e^{r(N+1-2n_0)}e^{s(N+1-2m_0)}(\hat{\tilde{p}}_{N,A}\,\delta_{n 1}+\hat{\tilde{p}}_{1,A}\,\delta_{n N}) \\
    &\quad -\epsilon \sin\phi \,(e^{-r(N-1)}\,e^{-s(N-1)}\,\hat{\tilde{x}}_{N,A}\,\delta_{n 1}-e^{r(N-1)}\,e^{s(N-1)}\,\hat{\tilde{x}}_{1,A}\,\delta_{n N}) ,\\
    \dot{\hat{\tilde{x}}}_{n,B} &= +\tilde{t}_1 \, \hat{\tilde{x}}_{n,A} - \tilde{t}_2 \, \hat{\tilde{x}}_{n+1,A} ,\\
    \dot{\hat{\tilde{p}}}_{n,A} &= -\tilde{t}_1 \, \hat{\tilde{p}}_{n,B} + \tilde{t}_2 \, \hat{\tilde{p}}_{n-1,B} -\frac{\kappa}{2}\,\hat{\tilde{p}}_{m,A}\,\delta_{n m} -  (\sqrt{2\kappa}\,\abs{\beta}\sin\theta + \sqrt{\kappa}\,\hat{P}^{\text{in}})\, e^{-r(m-n_0)}e^{-s(m-m_0)}\delta_{n m}\\
    &\quad -\epsilon \cos\phi \, e^{-r(N+1-2n_0)} e^{-s(N+1-2m_0)}(\hat{\tilde{x}}_{N,A}\,\delta_{n 1}+\hat{\tilde{x}}_{1,A}\,\delta_{n N}) \\
    &\quad +\epsilon \sin\phi \, (e^{-r(N-1)}\,e^{-s(N-1)}\,\hat{\tilde{p}}_{1,A}\,\delta_{n N}- e^{r(N-1)}\,e^{s(N-1)}\,\hat{\tilde{p}}_{N,A}\,\delta_{n 1}) ,\\
    \dot{\hat{\tilde{p}}}_{n,B} &= +\tilde{t}_1 \, \hat{\tilde{p}}_{n,A} - \tilde{t}_2 \, \hat{\tilde{p}}_{n+1,A} .
    \label{EomInOutQuadPert2Simp}
    \end{split}
\end{equation}
We Fourier transform Eq.~\eqref{EomInOutQuadPert2Simp} and define the zero-frequency susceptibility matrix $\chi[0,\epsilon] = (-\tilde{\mathds{H}}_{\text{NHSE}}(\epsilon))^{-1}$ where $\tilde{\mathds{H}}_{\text{NHSE}}(\epsilon) = \tilde{\mathds{H}}(\kappa)+\tilde{\mathds{H}}_{\text{pert,NHSE}}(\epsilon)$ with $\tilde{\mathds{H}}(\kappa)$ as defined in Eq.~\eqref{Hk} and
\begin{equation}
    \begin{split}
    \tilde{\mathds{H}}_{\text{pert,NHSE}}(\epsilon) =& -\epsilon \sin\phi \, e^{-(r+s)(N-1)}\ket{1,A}\bra{N,A} + \epsilon \cos\phi \, e^{r(N+1-2n_0)}e^{s(N+1-2m_0)}\ket{1,A}\bra{2N,A} \\
    &+\epsilon \sin\phi \, e^{(r+s)(N-1)} \ket{N,A}\bra{1,A} +\epsilon \cos\phi \, e^{r(N+1-2n_0)}e^{s(N+1-2m_0)}\ket{N,A}\bra{N\!+\!1,A}\\
    &-\epsilon \cos\phi \, e^{-r(N+1-2n_0)}e^{-s(N+1-2m_0)}\ket{N\!+\!1,A}\bra{N,A} -\epsilon \sin\phi \, e^{(r+s)(N-1)}\ket{N\!+\!1,A}\bra{2N,A}\\
    &-\epsilon \cos\phi \, e^{-r(N+1-2n_0)}e^{-s(N+1-2m_0)}\ket{2N,A}\bra{1,A} +\epsilon \sin\phi \, e^{-(r+s)(N-1)}\ket{2N,A}\bra{N\!+\!1,A} .
    \end{split}
\end{equation}
Then, the dynamics for the NHSE are given via
\begin{equation}
    \begin{split}
    \hat{\tilde{x}}_{n,A} =& \, [\tilde{\mathds{H}}_{\text{NHSE}}(\epsilon)]^{-1}_{2n-1,2m-1} \, (\sqrt{2\kappa}\abs{\beta}\cos\theta + \sqrt{\kappa}\hat{X}^{\text{in}})\,e^{r(m-n_0)} \, e^{s(m-m_0)} \,+ \\ 
    &+[\tilde{\mathds{H}}_{\text{NHSE}}(\epsilon)]^{-1}_{2n-1,2N+2m-2} \,(\sqrt{2\kappa}\,\abs{\beta}\sin\theta + \sqrt{\kappa}\,\hat{P}^{\text{in}})\,e^{-r(m-n_0)}\,e^{-s(m-m_0)} ,\\
    \hat{\tilde{x}}_{n,B} =& \, [\tilde{\mathds{H}}_{\text{NHSE}}(\epsilon)]^{-1}_{2n,2m-1} \, (\sqrt{2\kappa}\abs{\beta}\cos\theta + \sqrt{\kappa}\hat{X}^{\text{in}})\,e^{r(m-n_0)} \,e^{s(m-m_0)} \, + \\
    &+[\tilde{\mathds{H}}_{\text{NHSE}}(\epsilon)]^{-1}_{2n,2N+2m-2} \,(\sqrt{2\kappa}\,\abs{\beta}\sin\theta + \sqrt{\kappa}\,\hat{P}^{\text{in}})\,e^{-r(m-n_0)}\,e^{-s(m-m_0)} ,\\    
    \hat{\tilde{p}}_{n,A} =&  \, [\tilde{\mathds{H}}_{\text{NHSE}}(\epsilon)]^{-1}_{2N+2n-2,2m-1} \, (\sqrt{2\kappa}\abs{\beta}\cos\theta + \sqrt{\kappa}\hat{X}^{\text{in}})\,e^{r(m-n_0)} \, e^{s(m-m_0)} \,+\\
    &+[\tilde{\mathds{H}}_{\text{NHSE}}(\epsilon)]^{-1}_{2N+2n-2,2N+2m-2} \,(\sqrt{2\kappa}\,\abs{\beta}\sin\theta + \sqrt{\kappa}\,\hat{P}^{\text{in}})\,e^{-r(m-n_0)}\,e^{-s(m-m_0)} ,\\
    \hat{\tilde{p}}_{n,B} =& \, [\tilde{\mathds{H}}_{\text{NHSE}}(\epsilon)]^{-1}_{2N+2n-1,2m-1} \, (\sqrt{2\kappa}\abs{\beta}\cos\theta + \sqrt{\kappa}\hat{X}^{\text{in}})\,e^{r(m-n_0)} \, e^{s(m-m_0)} \, +\\
    &+ [\tilde{\mathds{H}}_{\text{NHSE}}(\epsilon)]^{-1}_{2N+2n-1,2N+2m-2} \,(\sqrt{2\kappa}\,\abs{\beta}\sin\theta + \sqrt{\kappa}\,\hat{P}^{\text{in}})\,e^{-r(m-n_0)}\,e^{-s(m-m_0)} ,
    \end{split}
    \label{eq:QuadPert2Simp}
\end{equation}
where $1\leq n \leq N$ with $\hat{\tilde{x}}_{N,B} =  \hat{\tilde{p}}_{N,B} \equiv 0$.

For an \textit{infinitesimal} $\epsilon$, we can proceed as we did with the on-site perturbation and compute the matrix elements of $[\tilde{\mathds{H}}_{\text{NHSE}}(\epsilon)]^{-1}$ at first order using Dyson's equation in terms of the matrix elements of $[\tilde{h}]^{-1}$ of the dynamical matrix (see Appendix~\ref{app:FirstHpert}). Then, applying the inverse of the squeezing transformation defined in Eq.~\eqref{SqueezTrans}, we can get $\hat{x}_{n,A}, \hat{x}_{n,B}, \hat{p}_{n,A}$ and $\hat{p}_{n,B}$ and get the relevant expectation values for the signal defined in Eq.~\eqref{SignalProt1}. For the NHSE perturbation we find
\begin{equation}
    \begin{split}
    \langle\hat{x}_{m,A} \rangle_\epsilon - \langle\hat{x}_{m,A} \rangle_0 &= -\epsilon \left( \frac{2}{\kappa}\right)^{\!\!2} \sin\phi \left( \frac{\gamma_2^2\!-\!t_2^2}{\gamma_1^2\!-\!t_1^2}\right)^{\!\!m} \left[  \left( \frac{\gamma_1\!+\!t_1}{\gamma_2\!-\!t_2}\right)^{\!\!N}  \!\!\left( \frac{\gamma_1\!-\!t_1}{\gamma_2\!+\!t_2}\right)  - \left( \frac{\gamma_1\!-\!t_1}{\gamma_2\!+\!t_2}\right)^{\!\!N} \!\! \left( \frac{\gamma_1\!+\!t_1}{\gamma_2\!-\!t_2}\right)  \right]  \sqrt{2\kappa}\,\abs{\beta}\,\cos\theta \\
    &-2\epsilon \left( \frac{2}{\kappa}\right)^{\!\!2} \cos\phi \,  \left( \frac{\gamma_1\!+\!t_1}{\gamma_2\!-\!t_2}\right)^{\!\!N-2m+1} \sqrt{2\kappa}\,\abs{\beta}\,\sin\theta ,
    \end{split}
\end{equation}
and
\begin{equation}
    \begin{split}
    \langle\hat{p}_{m,A} \rangle_\epsilon - \langle\hat{p}_{m,A} \rangle_0 &= 2\epsilon \left( \frac{2}{\kappa}\right)^{\!\!2} \cos\phi  \left( \frac{\gamma_1\!-\!t_1}{\gamma_2\!+\!t_2}\right)^{\!\!N-2m+1} \, \sqrt{2\kappa}\,\abs{\beta}\,\cos\theta \\
    &+\epsilon \left( \frac{2}{\kappa}\right)^{\!\!2} \sin\phi \left( \frac{\gamma_2^2\!-\!t_2^2}{\gamma_1^2\!-\!t_1^2}\right)^{\!\!m} \left[  \left( \frac{\gamma_1\!+\!t_1}{\gamma_2\!-\!t_2}\right)^{\!\!N}  \left( \frac{\gamma_1\!-\!t_1}{\gamma_2\!+\!t_2}\right) - \left( \frac{\gamma_1\!-\!t_1}{\gamma_2\!+\!t_2}\right)^{\!\!N} \!\!\left( \frac{\gamma_1\!+\!t_1}{\gamma_2\!-\!t_2}\right)  \right]  \sqrt{2\kappa}\,\abs{\beta}\,\sin\theta ,
    \end{split}
\end{equation}
so that the signal is given by
\begin{equation}
    \begin{split}
    \mathcal{S}_\tau(N,\epsilon) = 32 \tau \frac{\epsilon^2}{\kappa^2}\abs{\beta}^2 &\Bigg[ -\cos\varphi \, \Biggl( \sin\phi \left( \frac{\gamma_2^2-t_2^2}{\gamma_1^2-t_1^2}\right)^{\!\!m} \left[  \left( \frac{\gamma_1\!+\!t_1}{\gamma_2\!-\!t_2}\right)^{\!\!N} \!\! \left( \frac{\gamma_1\!-\!t_1}{\gamma_2\!+\!t_2}\right) - \left( \frac{\gamma_1\!-\!t_1}{\gamma_2\!+\!t_2}\right)^{\!\!N} \left( \frac{\gamma_1\!+\!t_1}{\gamma_2\!-\!t_2}\right)  \right] \cos\theta \, + \\
    &+2 \cos\phi \,  \left( \frac{\gamma_1\!+\!t_1}{\gamma_2\!-\!t_2}\right)^{\!\!N-2m+1} \, \sin\theta \Biggr ) + \sin\varphi \, \Biggl(  2 \cos\phi \,  \left( \frac{\gamma_1\!-\!t_1}{\gamma_2\!+\!t_2}\right)^{\!\!N-2m+1} \, \cos\theta \, +\\
    &+ \sin\phi \left( \frac{\gamma_2^2-t_2^2}{\gamma_1^2-t_1^2}\right)^m \left[  \left( \frac{\gamma_1+t_1}{\gamma_2-t_2}\right)^N  \left( \frac{\gamma_1\!-\!t_1}{\gamma_2\!+\!t_2}\right) - \left( \frac{\gamma_1\!-\!t_1}{\gamma_2\!+\!t_2}\right)^{\!\!N}\!\! \left( \frac{\gamma_1\!+\!t_1}{\gamma_2\!-\!t_2}\right)  \right] \, \sin\theta \Biggr) \Bigg]^2.
    \end{split}
    \label{SignalFirstNHSEFP}
\end{equation}
Finally, the noise and number of photons are the same as equations~\eqref{NoiseZerothFP} and~\eqref{PhotonsZerothFP}, respectively, since the zeroth order contributions are perturbation independent.

In \textit{regime I} and in the large N-limit we find that
\begin{equation}
    \overline{\text{SNR}}_\tau (N) \,  \approx  \frac{8\tau\epsilon^2}{\kappa\,(\frac{1}{2}\!+\!\bar{n}_{\text{th}})}  \, \frac{\mathds{A}^{\!I}(t_1,t_2,\gamma_1,\gamma_2,\phi,\theta,\varphi)}{\cos^2\theta \, \left( \frac{\gamma_2+t_2}{\gamma_1-t_1}\right)^{\!2(m-1)} + \sin^2\theta \left( \frac{\gamma_2-t_2}{\gamma_1+t_1}\right)^{\!2(m-1)}}  ,
\end{equation}
with 
\begin{equation}
    \begin{split}
    \mathds{A}^{\!I}(t_1,t_2,\gamma_1,\gamma_2,\phi,\theta,\varphi) &= \Bigg[\! -\cos\varphi \, \Biggl( \sin\phi \left( \frac{\gamma_2^2-t_2^2}{\gamma_1^2-t_1^2}\right)^{\!\!m} \left[  \left( \frac{\gamma_1\!+\!t_1}{\gamma_2\!-\!t_2}\right)^{\!\!N}  \!\!\left( \frac{\gamma_1\!-\!t_1}{\gamma_2\!+\!t_2}\right) - \left( \frac{\gamma_1\!-\!t_1}{\gamma_2\!+\!t_2}\right)^{\!\!N} \!\!\left( \frac{\gamma_1\!+\!t_1}{\gamma_2\!-\!t_2}\right)  \right] \cos\theta \, + \\
    &+2 \cos\phi \,  \left( \frac{\gamma_1\!+\!t_1}{\gamma_2\!-\!t_2}\right)^{\!\!N-2m+1} \, \sin\theta \Biggr ) + \sin\varphi \, \Biggl(  2 \cos\phi \,  \left( \frac{\gamma_1\!-\!t_1}{\gamma_2\!+\!t_2}\right)^{\!\!N-2m+1} \,\cos\theta \, +\\
    &+ \sin\phi \left( \frac{\gamma_2^2-t_2^2}{\gamma_1^2-t_1^2}\right)^{\!\!m} \left[  \left( \frac{\gamma_1\!+\!t_1}{\gamma_2\!-\!t_2}\right)^{\!\!N} \!\! \left( \frac{\gamma_1\!-\!t_1}{\gamma_2\!+\!t_2}\right) - \left( \frac{\gamma_1\!-\!t_1}{\gamma_2\!+\!t_2}\right)^{\!\!N} \!\!\left( \frac{\gamma_1\!+\!t_1}{\gamma_2\!-\!t_2}\right)  \right] \, \sin\theta \Biggr) \Bigg]^2 .
    \end{split}
    \label{eq:AI}
\end{equation}
We find exponential attenuation for all parameters with  $\overline{\text{SNR}}_\tau (N) \!\propto\! \left( \frac{\gamma_1+t_1}{\gamma_2-t_2}\right)^{\!2N}$ for $\varphi \in [0,\frac{\pi}{2})$ and $\forall \theta$ or $\varphi=\frac{\pi}{2}$ and $\theta\neq\{0,\pi\}$. A \textit{fine-tuned} exponential attenuation with rate $\overline{\text{SNR}}_\tau (N) \!\propto\! \left( \frac{\gamma_1-t_1}{\gamma_2+t_2}\right)^{\!2N}$ exists for $\varphi=\frac{\pi}{2}$ and $\theta=\{0,\pi\}$.

In \textit{regime II} and in the large N-limit we find that
\begin{equation}
    \overline{\text{SNR}}_\tau (N) \, \approx  \frac{8\tau\epsilon^2}{\kappa\,(\frac{1}{2}\!+\!\bar{n}_{\text{th}})}  \, \frac{\mathds{A}^{\!II} (t_1,t_2,\gamma_1,\gamma_2,\phi,\theta,\varphi)}{\cos^2\theta \, \left( \frac{\gamma_2+t_2}{\gamma_1-t_1}\right)^{\!2(m-1)} + \sin^2\theta \left( \frac{\gamma_1+t_1}{\gamma_2-t_2}\right)^{\!2(N-m)}},
\end{equation}
with
\begin{equation}
    \begin{split}
    \mathds{A}^{\!II} (t_1,t_2,\gamma_1,\gamma_2,\phi,\theta,\varphi) &= \Bigg[\! -\cos\varphi \, \Biggl( \sin\phi \left( \frac{\gamma_2^2-t_2^2}{\gamma_1^2-t_1^2}\right)^{\!\!m} \!\!  \left( \frac{\gamma_1\!+\!t_1}{\gamma_2\!-\!t_2}\right)^{\!\!N} \!\! \left( \frac{\gamma_1\!-\!t_1}{\gamma_2\!+\!t_2}\right)  \,\cos\theta \, + \\
    &+2 \cos\phi \,  \left( \frac{\gamma_1\!+\!t_1}{\gamma_2\!-\!t_2}\right)^{\!\!N-2m+1} \, \sin\theta \Biggr ) \, + \sin\varphi \, \Bigg( 2 \cos\phi \,  \left( \frac{\gamma_1\!-\!t_1}{\gamma_2\!+\!t_2}\right)^{\!\!N-2m+1} \, \cos\theta \, \\
    & +  \sin\phi \left( \frac{\gamma_2^2-t_2^2}{\gamma_1^2-t_1^2}\right)^m  \left( \frac{\gamma_1\!+\!t_1}{\gamma_2\!-\!t_2}\right)^{\!\!N}  \left( \frac{\gamma_1\!-\!t_1}{\gamma_2\!+\!t_2}\right)  \, \sin\theta \Bigg) \Bigg]^2 .
    \end{split}
\end{equation}
We find exponential amplification with $\overline{\text{SNR}}_\tau (N) \!\propto\! \left( \frac{\gamma_1+t_1}{\gamma_2-t_2}\right)^{\!2N}$ for $\varphi \in [0,\frac{\pi}{2})$ and $\theta = 0$. Additionally, it is found amplification with $\overline{\text{SNR}}_\tau (N) \!\propto\! \left( \frac{\gamma_2+t_2}{\gamma_1-t_1}\right)^{\!2N}$  if we drive the system at $m=N$ with $\varphi \in (0,\frac{\pi}{2}]$ and $\theta=\frac{\pi}{2}$. Remarkably, both cases are \textit{fine-tuned}, with similar amplifying paths for the BKS model reported in previous studies~\cite{BaoQiDong2022}. The main result in this regime is that
for most of the remaining parameter values, both the exponential amplification of the signal and the average photon number are proportional to $\left( \frac{\gamma_1+t_1}{\gamma_2-t_2}\right)^{\!2N}$, i.e., no true enhancement is observed. Lastly, one can find fine-tuned attenuation with $\overline{\text{SNR}}_\tau (N) \!\propto\! \left( \frac{\gamma_1-t_1}{\gamma_2+t_2}\right)^{\!2N}$ for $\varphi=\frac{\pi}{2}$ and $\theta = 0$ and with $\overline{\text{SNR}}_\tau (N) \!\propto\! \left( \frac{\gamma_2-t_2}{\gamma_1+t_1}\right)^{\!2N}$ if we drive the system at $m=N$ with $\varphi=0$ and $\theta = \frac{\pi}{2}$.

In \textit{regime III} and in the large N-limit we find that
\begin{equation}
    \overline{\text{SNR}}_\tau (N) \approx  \frac{8\tau\epsilon^2}{\kappa\,(\frac{1}{2}\!+\!\bar{n}_{\text{th}})}   \, \frac{\mathds{A}^{\!I} (t_1,t_2,\gamma_1,\gamma_2,\phi,\theta,\varphi)}{\cos^2\theta \, \left( \frac{\gamma_1-t_1}{\gamma_2+t_2}\right)^{\!2(N-m)} + \sin^2\theta \left( \frac{\gamma_1+t_1}{\gamma_2-t_2}\right)^{\!2(N-m)}} ,
\end{equation}
with $\mathds{A}^{\!I}(t_1,t_2,\gamma_1,\gamma_2,\phi,\theta,\varphi)$ given in Eq.~\eqref{eq:AI}. In this regime, we find fine-tuned exponential amplification with $\overline{\text{SNR}}_\tau (N) \!\propto\! \left( \frac{(\gamma_1+t_1)(\gamma_2+t_2)}{(\gamma_1-t_1)(\gamma_2-t_2)}\right)^{\!2N}$ for $\varphi \in [0,\frac{\pi}{2})$ and $\theta = \{0,\pi\}$. Besides, we can also find the system exhibiting exponential attenuation with $\overline{\text{SNR}}_\tau (N) \!\propto\! \left(\left( \frac{\gamma_2+t_2}{\gamma_1-t_1}\right)^{\!N} - \left( \frac{\gamma_2-t_2}{\gamma_1+t_1}\right)^{\!N}\right)^2$ if we drive the system at $m=N$ with $\varphi \in [0,\frac{\pi}{2}]$ and $\forall \theta$. For most of the remaining parameter values, both the exponential amplification of the signal and the average photon number are proportional to $\left( \frac{\gamma_1+t_1}{\gamma_2-t_2}\right)^{\!2N}$ (or proportional to$\left( \frac{\gamma_1-t_1}{\gamma_2+t_2}\right)^{\!2N}$ for $\varphi=\frac{\pi}{2}$ and $\theta = 0$), i.e., no true enhancement is observed.

\section{First order calculation of $[\tilde{\mathds{H}}(\epsilon)]^{-1}$}
\label{app:FirstHpert}
Here we compute the matrix elements $[\tilde{\mathds{H}}(\epsilon)]^{-1}$ which are needed to solve the system's dynamics. For an infinitesimal $\epsilon$, we will keep terms up to first order in $\epsilon$. This, using Dyson's equation, leads to the following expression:
\begin{equation}
    [\tilde{\mathds{H}}(\epsilon)]^{-1} = [\tilde{\mathds{H}}(\kappa)+\tilde{\mathds{H}}_{\text{pert}}(\epsilon)]^{-1} \simeq [\tilde{\mathds{H}}(\kappa)]^{-1} - [\tilde{\mathds{H}}(\kappa)]^{-1} \; \tilde{\mathds{H}}_{\text{pert}}(\epsilon) \; [\tilde{\mathds{H}}(\kappa)]^{-1} \,,
\end{equation}
where $[\tilde{\mathds{H}}(\kappa)]^{-1}$ is a block diagonal matrix with two identical blocks, each equal to $[\tilde{h}]^{-1}$ (see Appendix~\ref{app:tildeh}), and
\begin{equation}
     \begin{split}
     \tilde{\mathds{H}}_{\text{pert},N}(\epsilon) =& +\epsilon \Bigl(e^{2r(N-n_0)}e^{2s(N-m_0)}  \ket{2N\!-\!1}\bra{4N\!-\!2} - e^{-2r(N-n_0)}e^{-2s(N-m_0)} \ket{4N\!-\!2}\bra{2N\!-\!1} \Bigr) ,\\
      \tilde{\mathds{H}}_{\text{pert,NHSE}} (\epsilon) =& -\epsilon \sin\phi \, e^{-(r+s)(N-1)}\ket{1}\bra{2N\!-\!1} + \epsilon \cos\phi \, e^{r(N+1-2n_0)}e^{s(N+1-2m_0)}\ket{1}\bra{4N\!-\!2} \\
    &+\epsilon \sin\phi \, e^{(r+s)(N-1)}\ket{2N\!-\!1}\bra{1} +\epsilon \cos\phi \, e^{r(N+1-2n_0)}e^{s(N+1-2m_0)}\ket{2N\!-\!1}\bra{2N}\\
    &-\epsilon \cos\phi \, e^{-r(N+1-2n_0)}e^{-s(N+1-2m_0)}\ket{2N}\bra{2N\!-\!1} -\epsilon \sin\phi \, e^{(r+s)(N-1)}\ket{2N}\bra{4N\!-\!2}\\
    &-\epsilon \cos\phi \, e^{-r(N+1-2n_0)} e^{-s(N+1-2m_0)}\ket{4N-2}\bra{1} +\epsilon \sin\phi \, e^{-(r+s)(N-1)}\ket{4N\!-\!2}\bra{2N} .
     \end{split}
\end{equation}
for the on-site and the NHSE perturbations, respectively. 

The relevant terms needed to solve the dynamics of the system for the on-site perturbation are found to be
\begin{equation}
    \begin{split}
    [\tilde{\mathds{H}}_{N}(\epsilon)]^{-1}_{2m-1,2m-1} &= [\tilde{h}]^{-1}_{2m-1,2m-1} \,,  \\[5pt]
    [\tilde{\mathds{H}}_{N}(\epsilon)]^{-1}_{2m-1,2N+2m-2} &= -\epsilon \, e^{2r(N-n_0)}\,e^{2s(N-m_0)}\,[\tilde{h}]^{-1}_{2m-1,2N-1} \, [\tilde{h}]^{-1}_{2N-1,2m-1} \,, \\[5pt]
    [\tilde{\mathds{H}}_{N}(\epsilon)]^{-1}_{2N+2m-2,2m-1} &= \epsilon \, e^{-2r(N-n_0)}\,e^{-2s(N-m_0)}\,[\tilde{h}]^{-1}_{2N-1,2m-1}[\tilde{h}]^{-1}_{2m-1,2N-1}  \,, \\[5pt]
    [\tilde{\mathds{H}}_{N}(\epsilon)]^{-1}_{2N+2m-2,2N+2m-2} &= [\tilde{h}]^{-1}_{2m-1,2m-1}  .
    \end{split}
\end{equation}
For the NHSE perturbation we have
\begin{equation}
   \begin{split}
    [\tilde{\mathds{H}}_{\text{NHSE}}(\epsilon)]^{-1}_{2m-1,2m-1} &= [\tilde{h}]^{-1}_{2m-1,2m-1} + \epsilon \sin\phi \, e^{-(r+s)(N-1)}[\tilde{h}]^{-1}_{2m-1,1}\, [\tilde{h}]^{-1}_{2N-1,2m-1} \\
    &-\epsilon\sin\phi \, e^{(r+s)(N-1)} [\tilde{h}]^{-1}_{2m-1,2N-1}\,[\tilde{h}]^{-1}_{1,2m-1}  \,,\\[5pt]
    [\tilde{\mathds{H}}_{\text{NHSE}}(\epsilon)]^{-1}_{2m-1,2N+2m-2} &= -\epsilon \cos\phi \, e^{r(N+1-2n_0)}e^{s(N+1-2m_0)}\,[\tilde{h}]^{-1}_{2m-1,1} \, [\tilde{h}]^{-1}_{2N-1,2m-1} \\
    &-\epsilon \cos\phi \, e^{r(N+1-2n_0)}e^{s(N+1-2m_0)}\,[\tilde{h}]^{-1}_{2m-1,2N-1} \, [\tilde{h}]^{-1}_{1,2m-1} \,,\\[5pt]
    [\tilde{\mathds{H}}_{\text{NHSE}}(\epsilon)]^{-1}_{2N+2m-2,2m-1} &= \epsilon \cos\phi \, e^{-r(N+1-2n_0)}e^{-s(N+1-2m_0)}\,[\tilde{h}]^{-1}_{2m-1,1}\,[\tilde{h}]^{-1}_{2N-1,2m-1}   \\
    &+\epsilon \cos\phi \, e^{-r(N+1-2n_0)}e^{-s(N+1-2m_0)}\,[\tilde{h}]^{-1}_{2m-1,2N-1}\,[\tilde{h}]^{-1}_{1,2m-1} \,,\\[5pt]
    [\tilde{\mathds{H}}_{\text{NHSE}}(\epsilon)]^{-1}_{2N+2m-2,2N+2m-2} &= [\tilde{h}]^{-1}_{2m-1,2m-1} + \epsilon \sin\phi \, e^{(r+s)(N-1)}[\tilde{h}]^{-1}_{2m-1,1} \,[\tilde{h}]^{-1}_{2N-1,2m-1} \\
    &-\epsilon\sin\phi \, e^{-(r+s)(N-1)} [\tilde{h}]^{-1}_{2m-1,2N-1}\,[\tilde{h}]^{-1}_{1,2m-1} \,.
  \end{split}
\end{equation}
Finally, to express the matrix elements $[\tilde{\mathds{H}}(\epsilon)]^{-1}$ in terms of the system parameters, we must determine the matrix elements of the inverse matrix $[\tilde{h}]^{-1}$, which are provided in Appendix~\ref{app:tildeh}.

\section{Matrix elements of $[\tilde{h}]^{-1}$.}
\label{app:tildeh}
For the (anti-Hermitian) matrix
\begin{equation}
   \begin{split}
    \tilde{h} = -\frac{\kappa}{2} \ket{m,A}\bra{m,A} &+ \sum_{n=1}^{N-1} \Bigl[ \tilde{t}_1 \Big( -\ket{n,A}\bra{n,B} + \ket{n\!-\!1,B}\bra{n,A} \Big) \Bigr] + \\
    &+\sum_{n=2}^{N} \Bigl[ \tilde{t}_2 \Big( \ket{n,A}\bra{n\!-\!1,B} -  \ket{n\!-\!1,B}\bra{n,A} \Big) \Bigr] \,,
    \label{eq:matrixhtilde}
    \end{split}
\end{equation}
the matrix elements relevant for solving the system's dynamics are found to be
\begin{align}
     [\tilde{h}]^{ -1}_{2n, 1} &= -\frac{1}{\tilde{t}_1}\left(\frac{\tilde{t}_2}{\tilde{t}_1}\right)^{\!\!n-1} ,&    &1 \leq n \leq m-1; &  [\tilde{h}]^{-1}_{2n, 2N-1} &= 0 \,, & &1 \leq n \leq m\!-\!1 ; \notag \\
     [\tilde{h}]^{-1}_{2n, 1} &= 0 \, , &  &m \leq n \leq N\!-\!1 ; & [\tilde{h}]^{-1}_{2n, 2N-1} &= \frac{1}{\tilde{t}_2}\left(\frac{\tilde{t}_1}{\tilde{t}_2}\right)^{\!\!N-n-1} , &  &m \leq n \leq N\!-\!1 ; \\
     [\tilde{h}]^{-1}_{2n-1, 1} &= -\frac{2}{\kappa} \left(\frac{\tilde{t}_2}{\tilde{t}_1}\right)^{\!\!2m-n-1} , &  &1 \leq n \leq N ; &  [\tilde{h}]^{-1}_{2n-1, 2N-1} &= -\frac{2}{\kappa} \left(\frac{\tilde{t}_1}{\tilde{t}_2}\right)^{\!\!N+n-2m} , &  &1 \leq n \leq N ; \notag\\
     [\tilde{h}]^{-1}_{2n, 2m-1} &= 0 \,,&   &1 \leq n \leq N\!-\!1 ; & & && \notag\\
     [\tilde{h}]^{-1}_{2n-1, 2m-1} &= -\frac{2}{\kappa} \left(\frac{\tilde{t}_2}{\tilde{t}_1}\right)^{\!\!m-n} ,&   &1 \leq n \leq N .& & &&\notag
\end{align}

\section{Relation between $h^X$, $h^P$ and $[\tilde{h}]$.}
\label{app:hXPtildeh}

With the squeezing transformation in Eq.~\eqref{SqueezTrans} we can relate the matrix elements of $h^X,h^P$ with the matrix elements of $\tilde{h}$. By defining $T= \text{diag}(e^{-r},e^{-2r},e^{-2r},e^{-3r},\dots,e^{-N r})\cdot \text{diag}(e^{-s},e^{-s},e^{-2s},e^{-2s},\dots,e^{-N s})$ we have
\begin{equation}
    \begin{split}
    h^X &= T \, \tilde{h} \, T^{-1} , \\
    h^P &= T^{-1} \, \tilde{h} \, T .
    \end{split}
\end{equation}
If both $\{i,j\}$ are even or odd then
\begin{equation}
    \begin{split}
        (h^X)^{^{-1}}_{i,j} &= (\tilde{h})^{^{-1}}_{i,j} \, e^{\frac{r+s}{2}(j-i)} , \\
        (h^P)^{^{-1}}_{i,j} &= (\tilde{h})^{^{-1}}_{i,j} \, e^{\frac{r+s}{2}(i-j)} .
    \end{split}
\end{equation}
If $i$ is odd and $j$ is even
\begin{equation}
    \begin{split}
        (h^X)^{^{-1}}_{i,j} &= (\tilde{h})^{^{-1}}_{i,j} \, e^{\frac{r+s}{2}(j-i)}e^{\frac{r-s}{2}} , \\
        (h^P)^{^{-1}}_{i,j} &= (\tilde{h})^{^{-1}}_{i,j} \, e^{\frac{r+s}{2}(i-j)}e^{\frac{s-r}{2}} .
    \end{split}
\end{equation}
If $i$ is even and $j$ is odd
\begin{equation}
    \begin{split}
        (h^X)^{^{-1}}_{i,j} &= (\tilde{h})^{^{-1}}_{i,j} \, e^{\frac{r+s}{2}(j-i)}e^{\frac{s-r}{2}} , \\
        (h^P)^{^{-1}}_{i,j} &= (\tilde{h})^{^{-1}}_{i,j} \, e^{\frac{r+s}{2}(i-j)}e^{\frac{r-s}{2}} .
    \end{split}
\end{equation}

\section{Dynamics Beyond the Linear Response}
\label{app:BeyondLinearResponse}
Here, we no longer assume the perturbation to be infinitesimally small, so contributions from all orders are considered. Specifically, we now need to account for noise beyond the zeroth order.  The noise at all orders can be expressed as
\begin{equation}
    \begin{split}
        \mathcal{N}_\tau (N,\epsilon_0) = \left(  \bar{n}_{\text{th}} \!+\! \frac{1}{2} \right) &\Bigl[ \, \cos^2\varphi \, \Bigl\{( 1+\kappa \, [\mathds{H}(\epsilon_0)]^{-1}_{2m-1,2m-1} )^2 + \kappa^2 ([\mathds{H}(\epsilon_0)]^{-1}_{2m-1,2N+2m-2})^2   \Bigr\}  \, +\\
        &+ \sin^2 \, \varphi \, \Bigl\{ ( 1+\kappa \, [\mathds{H}(\epsilon_0)]^{-1}_{2N+2m-2,2N+2m-2} )^2 + \kappa^2 ([\mathds{H}(\epsilon_0)]^{-1}_{2N+2m-2,2m-1})^2  \Bigr\} \, + \\
        &+ \sin\varphi \, \cos\varphi \, \Bigl\{ ( 1+\kappa \, [\mathds{H}(\epsilon_0)]^{-1}_{2m-1,2m-1} ) \, \kappa \, [\mathds{H}(\epsilon_0)]^{-1}_{2N+2m-2,2m-1}  \,  +\\
        &+ \kappa \, [\mathds{H}(\epsilon_0)]^{-1}_{2m-1,2N+2m-2}   \, ( 1+\kappa \, [\mathds{H}(\epsilon_0)]^{-1}_{2N+2m-2,2N+2m-2} ) \Bigr\}   \Bigr] ,
    \end{split}
\end{equation}
which depends on the perturbation considered.

\subsection{Solving the dynamics for the on-site perturbation.}
\label{app:DynamicsAllOnsitePert}
We fix the following parameters: the drive position $m=1$ and the measuring angle $\varphi=0$ (i.e., we measure the $x$-quadrature). For the on-site perturbation $\hat{V}_{N}$ we consider a drive with $\theta = \frac{\pi}{2}$, for simplicity. We then substitute the matrix elements of $[\tilde{\mathds{H}_{N}}(\epsilon)]^{-1}$, computed to all orders (see Appendix~\ref{app:AllHpert}), into Eqs.~\eqref{eq:QuadPert1Simp}. For our chosen parameters, the signal at all orders is found to be
\begin{equation}
    \mathcal{S}_\tau(N,\epsilon_0) = \kappa \tau \, \big| \langle\hat{x}_{1,A} \rangle_\epsilon - \langle\hat{x}_{1,A} \rangle_0 \big|^2 =  \frac{2 \, \tau \, \epsilon_0^2 \,\kappa^2 \abs{\beta}^2}{\left[ \epsilon_0^2 + \left(\frac{\kappa}{2}\right)^2 \, \left( \frac{\gamma_2^2-t_2^2}{\gamma_1^2-t_1^2}\right)^{\!2(N-1)}   \right]^2} \left( \frac{\gamma_2\!+\!t_2}{\gamma_1\!-\!t_1}\right)^{\!\!4(N-1)} .
     \label{SignalAllOnsiteFP}
\end{equation}
The noise at all orders, which will also be amplified as  $N$ grows, is given by
\begin{equation}
   \begin{split}
   \!\!\!\frac{\mathcal{N}_\tau (N,0)+\mathcal{N}_\tau (N,\epsilon_0)}{2} &= 
  \frac{1}{2}\left(  \bar{n}_{\text{th}} \! + \!\frac{1}{2} \right) \left( 1 + \left[ ( 1+\kappa \, [\mathds{H}_{N}(\epsilon_0)]^{-1}_{1,1} )^2 + \kappa^2 ([\mathds{H}_{N}(\epsilon_0)]^{-1}_{1,2N})^2  \right]  \right)  \\
    &=\frac{(2\,\bar{n}_{\text{th}}+1)}{2\left[ \epsilon_0^2 + \left(\frac{\kappa}{2}\right)^2  \left( \frac{\gamma_2^2-t_2^2}{\gamma_1^2-t_1^2}\right)^{2(N-1)} \right]^2} \Bigg\{ \!\! \left(\frac{\kappa}{2}\right)^{\!\!4}   \left(\frac{\gamma_2^2\!-\!t_2^2}{\gamma_1^2\!-\!t_1^2}\right)^{\!\!4(N-1)} \!+ \frac{\epsilon_0^2\,\kappa^2}{2} \left( \frac{\gamma_2\!+\!t_2}{\gamma_1\!-\!t_1}  \right)^{\!\!4(N-1)} + \epsilon_0^4 \Bigg\}.
    \end{split}
    \label{NoiseAllOnsiteFP}
\end{equation}
Note that, for our choice of parameters, the signal in Eq.~\eqref{SignalAllOnsiteFP} reduces to Eq.~\eqref{SignalFirstOnsiteFP} for $\epsilon_0 \ll 1$ and  the noise in Eq.~\eqref{NoiseAllOnsiteFP} reduces to Eq.~\eqref{NoiseZerothFP} for $\epsilon_0 = 0$, as expected. The signal-to-noise ratio defined in Eq.~\eqref{eq:SNRAllDef} reads then as
\begin{equation}
    \text{SNR}_\tau (N,\epsilon_0) = \frac{1}{(2\,\bar{n}_{\text{th}}\!+\!1)}\frac{8\tau\, \epsilon_0^{2}\,\kappa^2\,\abs{\beta}^2 \, \left( \frac{\gamma_2+t_2}{\gamma_1-t_1}\right)^{\!\!4(N-1)}}{\left[2\left(\frac{\kappa}{2}\right)^4\left(\frac{\gamma_2^2-t_2^2}{\gamma_1^2-t_1^2}\right)^{\!4(N-1)}  +  \epsilon_0^2\,\kappa^2\left( \frac{\gamma_2+t_2}{\gamma_1-t_1}\right)^{\!4(N-1)} + 2\,\epsilon_0^{4} \right] } .
\end{equation}
The signal in Eq.~\eqref{SignalAllOnsiteFP} scales as $\mathcal{S}_\tau(N,\epsilon_0) \!\propto\! \left( \frac{\gamma_1+t_1}{\gamma_2-t_2}\right)^{\!4N}$. Initially, the noise in Eq.~\eqref{SignalFirstOnsiteFP} is dominated by the term proportional to $\left(\frac{\kappa}{2}\right)^4$, keeping the noise at the vacuum level, $\left(  \bar{n}_{\text{th}} \!+\! \frac{1}{2} \right)$. However, beyond a certain point given by condition in Eq.~\eqref{eq:conditionOnsite} in the main text, the term proportional to $\epsilon_0^2 \kappa^2$ takes over, causing the noise to scale as $\mathcal{N}_\tau (N,\epsilon_0) \!\propto\!\left( \frac{\gamma_1+t_1}{\gamma_2-t_2}\right)^{\!4N}$, matching the signal's scaling. As a result, the signal-to-noise ratio, $\text{SNR}_\tau (N,\epsilon_0)$, saturates at the value $\frac{8\tau\abs{\beta}^2}{(2\,\bar{n}_{\text{th}}+1)}$. In the BKC limit, we set $t_2=t_1, \gamma_2=\gamma_1$, and $\bar{N}=2N-1$, where $\bar{N}$ represents the number of unit cells in the BKC. Thus, we find that
\begin{equation}
    \text{SNR}_\tau (N,\epsilon_0) = \frac{1}{(2\,\bar{n}_{\text{th}}\!+\!1)}\frac{8\tau\,\left(\frac{\epsilon_0\,\kappa}{\left(\frac{\kappa}{2}\right)^{\!2}+\epsilon_0^2}\right)^{\!\!2}\, \,\abs{\beta}^2 \, \left(\frac{\gamma_1+t_1}{\gamma_1-t_1} \right)^{\!\!2(\bar{N}-1)}}{1 + \left(\frac{\left(\frac{\kappa}{2}\right)^{\!2}-\epsilon_0^2}{\left(\frac{\kappa}{2}\right)^{\!2}+\epsilon_0^2}\right)^{\!\!2} +  \left(\frac{\epsilon_0\,\kappa}{\left(\frac{\kappa}{2}\right)^{\!2}+\epsilon_0^2}\right)^{\!\!2} \, \left(\frac{\gamma_1+t_1}{\gamma_1-t_1} \right)^{\!\!2(\bar{N}-1)}  } ,
\end{equation}
which agrees with the result presented in Ref.~\onlinecite{McDonald2020}. Finally, the number of photons takes the following expression
\begin{equation}
    \begin{split}
        \bar{n}_{\text{tot}} (N,\epsilon_0) &= \frac{\kappa \, \abs{\beta}^2}{\left[ \epsilon_0^2 + \left( \frac{\kappa}{2}\right)^{\!2}\, \left(\frac{\gamma_2^2-t_2^2}{\gamma_1^2-t_1^2)}\right)^{\!2(N-1)} \right]^2} \times \Biggl( \left(\frac{\kappa}{2}\right)^{\!\!2} \, \left(\frac{\gamma_2^2-t_2^2}{\gamma_1^2-t_1^2}\right)^{\!\!4(N-1)} \, \frac{1-\left( \frac{\gamma_1+t_1}{\gamma_2-t_2}\right)^{\!2N}}{1-\left( \frac{\gamma_1+t_1}{\gamma_2-t_2}\right)^{\!2}} + \\ 
        &+ \epsilon_0^2 \, \left( \frac{\gamma_2\!+\!t_2}{\gamma_1\!-\!t_1}\right)^{\!\!4(N-1)}  \, \frac{1-\left( \frac{\gamma_1-t_1}{\gamma_2+t_2}\right)^{\!2N}}{1-\left( \frac{\gamma_1-t_1}{\gamma_2+t_2}\right)^{\!2}} \, + \frac{\epsilon_0^4}{(\gamma_1\!-\!t_1)^2} \, \left(\frac{\gamma_2\!+\!t_2}{\gamma_2\!-\!t_2}\right) \, \frac{1-\left( \frac{\gamma_2+t_2}{\gamma_1-t_1}\right)^{\!2(N-1)}}{1-\left( \frac{\gamma_2+t_2}{\gamma_1-t_1}\right)^{\!2}} \\
        &+ \left( \frac{\kappa}{2}\right)^{\!\!2} \, \frac{\epsilon_0^2}{(\gamma_1\!+\!t_1)^2}\, 
         \left(\frac{\gamma_2\!+\!t_2}{\gamma_1\!-\!t_1}\right)^{\!\!4(N-1)}   \, \left(\frac{\gamma_2\!-\!t_2}{\gamma_2\!+\!t_2}\right)\,\frac{1-\left( \frac{\gamma_2-t_2}{\gamma_1+t_1}\right)^{\!2(N-1)}}{1-\left( \frac{\gamma_2-t_2}{\gamma_1+t_1}\right)^{\!2}} \, \Biggr) . 
    \end{split}
    \label{PhotonsAllOnsiteFP}
\end{equation}
Note that, for our choice of parameters, the expression in Eq.~\eqref{PhotonsAllOnsiteFP} reduces to Eq.~\eqref{PhotonsZerothFP} when $\epsilon_0 = 0$.

\subsection{Solving the dynamics for the NHSE perturbation.}
\label{app:DynamicsAllNHSEPert}
We fix the following parameters: the driving phase $\theta=\frac{\pi}{4}$, the NHSE perturbation phase $\phi=\frac{\pi}{2}$ and the measuring angle $\varphi=0$ (i.e., we measure the $x$-quadrature). We do not fix the position of the drive $m$. Substituting the matrix elements of $[\tilde{\mathds{H}}_{\text{NHSE}}(\epsilon)]^{-1}$, computed to all orders (see Appendix~\ref{app:AllHpert}), into Eqs.~\eqref{eq:QuadPert2Simp}, we find that the signal at all orders for our chosen parameters is given by
\begin{equation}
    \mathcal{S}_\tau(N,\epsilon_0) = \kappa \tau \, \big| \langle\hat{x}_{m,A} \rangle_\epsilon - \langle\hat{x}_{m,A} \rangle_0 \big|^2 = \frac{4 \, \tau \, \epsilon_0^2 \, \abs{\beta}^2}{\left[ \frac{\kappa}{2} - \epsilon_0 \, \mathcal{T} \right]^2} \mathcal{T}^2 ,
    \label{SignalAllNHSEFP}
\end{equation}
with
\begin{equation}
    \mathcal{T} = \left\{ \left( \frac{\gamma_2\!+\!t_2}{\gamma_1\!-\!t_1}\right)^{\!\!m-1} \left( \frac{\gamma_1\!+\!t_1}{\gamma_2\!-\!t_2}\right)^{\!\!N-m}- \left( \frac{\gamma_2\!-\!t_2}{\gamma_1\!+\!t_1}\right)^{\!\!m-1} \left( \frac{\gamma_1\!-\!t_1}{\gamma_2\!+\!t_2}\right)^{\!\!N-m}  \right\} .
\end{equation}
In \textit{regime II}, which is the case in which we can have exponential enhancement, this can be written as
\begin{equation}
    \mathcal{S}_\tau(N,\epsilon_0) \approx \frac{4 \, \tau \, \epsilon_0^2 \, \abs{\beta}^2}{\left[ \frac{\kappa}{2} - \epsilon_0 \, \left( \frac{\gamma_2+t_2}{\gamma_1-t_1}\right)^{\!m-1} \left( \frac{\gamma_1+t_1}{\gamma_2-t_2}\right)^{\!N-m} \right]^2}  \left( \frac{\gamma_2\!+\!t_2}{\gamma_1\!-\!t_1}\right)^{\!\!2(m-1)} \left( \frac{\gamma_1\!+\!t_1}{\gamma_2\!-\!t_2}\right)^{\!\!2(N-m)} .
\end{equation}
Likewise, the noise at all orders can be computed to be
\begin{equation}
    \begin{split}
    \!\!\!\!\frac{\mathcal{N}_\tau (N,0)+\mathcal{N}_\tau (N,\epsilon_0)}{2} &= \frac{1}{2}\left(  \bar{n}_{\text{th}} \!+\! \frac{1}{2} \right) \left( 1 \! + \!\left[  ( 1\!+\!\kappa \, [\mathds{H}_{\text{NHSE}}(\epsilon_0)]^{-1}_{2m-1,2m-1} )^2 + \kappa^2 ([\mathds{H}_{\text{NHSE}}(\epsilon_0)]^{-1}_{2m-1,2N+2m-2})^2  \right] \right)\\
    &= \frac{1}{2}\,\frac{(2\,\bar{n}_{\text{th}} \!+\! 1)}{\left[\frac{\kappa}{2} - \epsilon_0 \,  \mathcal{T}\right]^2}  \left[ \left(\frac{\kappa}{2}\right)^{\!\!2} + \epsilon_0^2 \,  \mathcal{T}^2   \right] .
    \end{split}
    \label{NoiseAllNHSEFP}
\end{equation}
which for \textit{regime II} this is
\begin{equation}
     \!\!\!\!\frac{\mathcal{N}_\tau (N,0)+\mathcal{N}_\tau (N,\epsilon_0)}{2} \approx 
      \frac{1}{2}\,\frac{(2\,\bar{n}_{\text{th}} \! + \!1)}{\left[\frac{\kappa}{2} - \epsilon_0  \left( \frac{\gamma_2+t_2}{\gamma_1-t_1}\right)^{\!m-1} \left( \frac{\gamma_1+t_1}{\gamma_2-t_2}\right)^{\!N-m}\right]^2}  \!\! \left[ \left(\frac{\kappa}{2}\right)^{\!\!2} + \epsilon_0^2 \,  \left( \frac{\gamma_2\!+\!t_2}{\gamma_1\!-\!t_1}\right)^{\!\!2(m-1)} \left( \frac{\gamma_1\!+\!t_1}{\gamma_2\!-\!t_2}\right)^{\!\!2(N-m)}   \right] . 
\end{equation}
Note that, for our choice of parameters, the signal in Eq.~\eqref{SignalAllNHSEFP} reduces to Eq.~\eqref{SignalFirstNHSEFP} for $\epsilon_0 \ll 1$ and  the noise in Eq.~\eqref{NoiseAllNHSEFP} reduces to Eq.~\eqref{NoiseZerothFP} for $\epsilon_0 = 0$, as expected. The signal-to-noise ratio defined in Eq.~\eqref{eq:SNRAllDef} is then given by
\begin{equation}
    \text{SNR}_\tau (N,\epsilon_0) = \frac{1}{(2\,\bar{n}_{\text{th}}\!+\!1)}\frac{8\tau\, \epsilon_0^{2}\,\abs{\beta}^2}{\left(\frac{\kappa}{2}\right)^{\!2} + \epsilon_0^2 \left( \frac{\gamma_2+t_2}{\gamma_1-t_1}\right)^{\!2(m-1)} \left( \frac{\gamma_1+t_1}{\gamma_2-t_2}\right)^{\!2(N-m)} } \left( \frac{\gamma_2\!+\!t_2}{\gamma_1\!-\!t_1}\right)^{\!\!2(m-1)} \left( \frac{\gamma_1\!+\!t_1}{\gamma_2\!-\!t_2}\right)^{\!\!2(N-m)} .
\end{equation}
We observe that the signal increases exponentially with  $N$ as long as the term proportional to $\left(\frac{\kappa}{2}\right)^2$ dominates the denominator in in Eq.~\eqref{SignalAllNHSEFP}. However, once the term proportional to $\epsilon_0$ becomes the dominant one, the signal saturates. A different behavior is seen in the noise in Eq.~\eqref{SignalFirstNHSEFP}, where it mostly remains at the vacuum level, $\left(  \bar{n}_{\text{th}} \!+\! \frac{1}{2} \right)$. Initially, the terms proportional to $\left(\frac{\kappa}{2}\right)^2$ dominate both the numerator and denominator, effectively canceling each other out. Once the condition in Eq.~\eqref{eq:conditionNHSE} (main text) is met, the terms proportional to $\epsilon_0$ take over in both the numerator and denominator, canceling out again. As a result, the signal-to-noise ratio, $\text{SNR}_\tau (N,\epsilon_0)$, saturates at the value $\frac{8\tau\abs{\beta}^2}{(2\,\bar{n}_{\text{th}}+1)}$. Finally, the number of photons takes the following expression
\begin{equation}
    \begin{split}
        \bar{n}_{\text{tot}} (N,\epsilon_0) &\approx   \frac{\kappa}{2}  \abs{\beta}^2 \frac{1}{\left( \frac{\kappa}{2} - \epsilon_0 \left( \frac{\gamma_2+t_2}{\gamma_1-t_1}\right)^{\!m-1} \left( \frac{\gamma_1+t_1}{\gamma_2-t_2}\right)^{\!N-m} \right)^2} \times \\
        &\qquad \qquad \left[ \left( \frac{\gamma_2\!+\!t_2}{\gamma_1\!-\!t_1}\right)^{\!\!2(m-1)}+ \frac{\epsilon_0^2}{(\gamma_1\!+\!t_1)^2} \left( \frac{\gamma_1\!+\!t_1}{\gamma_2\!-\!t_2}\right)^{\!\!2(N-m)} \left( \frac{\gamma_2\!+\!t_2}{\gamma_1\!-\!t_1}\right)^{\!\!2(m-1)} \right] \\
        &+ \frac{\kappa}{2}  \abs{\beta}^2 \frac{1}{\left( \frac{\kappa}{2} + \epsilon_0 \left( \frac{\gamma_2\!+\!t_2}{\gamma_1\!-\!t_1}\right)^{\!m-1} \left( \frac{\gamma_1\!+\!t_1}{\gamma_2\!-\!t_2}\right)^{\!N-m} \right)^2} \times \\
        &\qquad \qquad \left[ \left( \frac{\gamma_2\!-\!t_2}{\gamma_1\!+\!t_1}\right)^{\!\!2(m-1)}\frac{1-\left( \frac{\gamma_1+t_1}{\gamma_2-t_2}\right)^{\!2N}}{1-\left( \frac{\gamma_1+t_1}{\gamma_2-t_2}\right)^{\!2}}  + \frac{\epsilon_0^2}{(\gamma_1\!-\!t_1)^2} \left( \frac{\gamma_1\!+\!t_1}{\gamma_2\!-\!t_2}\right)^{\!\!2(N-m)} \left( \frac{\gamma_2\!+\!t_2}{\gamma_1\!-\!t_1}\right)^{\!\!2(m-2)} \right]   .
    \end{split}
    \label{PhotonsAllNHSEFP}
\end{equation}
Note that, for our choice of parameters, the expression in Eq.~\eqref{PhotonsAllNHSEFP} reduces to Eq.~\eqref{PhotonsZerothFP} when $\epsilon_0 = 0$.

\section{All orders calculation of $[\tilde{\mathds{H}}(\epsilon)]^{-1}$}
\label{app:AllHpert}
Here we compute the matrix elements $[\tilde{\mathds{H}}(\epsilon)]^{-1}$, which are necessary to solve the system's dynamics beyond linear response for a finite $\epsilon_0$. Specifically, we need to compute
\begin{equation}
    [\tilde{\mathds{H}}(\epsilon_0)]^{-1} = [\tilde{\mathds{H}}(\kappa)+\tilde{\mathds{H}}_{\text{pert}}(\epsilon_0)]^{-1} ,
\end{equation}
where $[\tilde{\mathds{H}}(\kappa)]^{-1}$ is a block diagonal matrix with two equal blocks $[\tilde{h}]^{-1}$ (see Appendix~\ref{app:tildeh}) and the perturbation matrices $\tilde{\mathds{H}}_{\text{pert}}(\epsilon_0)$ for the on-site and the NHSE perturbations are given by
\begin{equation}
     \begin{split}
     \tilde{\mathds{H}}_{\text{pert},N}(\epsilon_0) =& + \epsilon_0 \Bigl(e^{2r(N-n_0)}e^{2s(N-m_0)}-  \ket{2N\!-\!1}\bra{4N\!-\!2} - e^{-2r(N-n_0)} e^{-2s(N-m_0)} \ket{4N\!-\!2}\bra{2N\!-\!1} \Bigr) ,\\
     \tilde{\mathds{H}}_{\text{pert,NHSE}}(\epsilon_0) =& -\epsilon_0 \, e^{-(r+s)(N-1)} \ket{1}\bra{2N\!-\!1} + \epsilon_0 \, e^{(r+s)(N-1)}\ket{2N\!-\!1}\bra{1} \\
    & -\epsilon_0  \, e^{(r+s)(N-1)}\ket{2N}\bra{4N\!-\!2} +\epsilon_0 \, e^{-(r+s)(N-1)}\ket{4N\!-\!2}\bra{2N} .
     \end{split}
\end{equation}

The relevant matrix elements for the on-site perturbation, computed to all orders, are given by
\begin{equation}
    \begin{alignedat}{2}
        [\tilde{\mathds{H}}_{N}(\epsilon_0)]^{-1}_{2n-1,1} &= - \left( \frac{\tilde{t}_1}{\tilde{t}_2}\right)^{\!\!n-1}\frac{ \kappa }{2 \left[ \, \epsilon_0^2  + \left(\frac{\kappa}{2}\right)^{\!2} \, \left( \frac{\tilde{t}_2}{\tilde{t}_1} \right)^{\!4(N-1)} \right] } \left( \frac{\tilde{t}_2}{\tilde{t}_1}  \right)^{\!\!4(N-1)} , \qquad &&n=1,\dots,N  ;\\
        [\tilde{\mathds{H}}_{N}(\epsilon_0)]^{-1}_{2n-1,2N} &= -\left( \frac{\tilde{t}_2}{\tilde{t}_1}\right)^{\!\!N-n} \frac{\epsilon_0\,  e^{2r(N-n_0)}e^{2s(N-m_0)}}{ \left[\epsilon_0^2  + \left(\frac{\kappa}{2}\right)^{\!2}  \, \left( \frac{\tilde{t}_2}{\tilde{t}_1} \right)^{\!4(N-1)} \right]} \, \left( \frac{\tilde{t}_2}{\tilde{t}_1}  \right)^{\!\!N-1}  , \qquad &&n=1,\dots,N  ;\\
        [\tilde{\mathds{H}}_{N}(\epsilon_0)]^{-1}_{2n,2N} &= \left( \frac{\tilde{t}_2}{\tilde{t}_1}\right)^{\!\!n-1} \frac{\kappa}{2\,\tilde{t}_1}\frac{\epsilon_0\,e^{2r(N-n_0)}e^{2s(N-m_0)} }{\left[\epsilon_0^2  + \left(\frac{\kappa}{2}\right)^{\!2}  \, \left( \frac{\tilde{t}_2}{\tilde{t}_1} \right)^{\!4(N-1)} \right]} \, \left( \frac{\tilde{t}_2}{\tilde{t}_1}  \right)^{\!\!2(N-1)}  , \qquad &&n=1,\dots,N\!-\!1  ;\\
        [\tilde{\mathds{H}}_{N}(\epsilon_0)]^{-1}_{2N+2n-2,2N} &= -\left( \frac{\tilde{t}_2}{\tilde{t}_1}\right)^{\!\!N-n} \frac{\kappa }{2 \left[ \, \epsilon_0^2  + \left(\frac{\kappa}{2}\right)^{\!2}  \, \left( \frac{\tilde{t}_2}{\tilde{t}_1} \right)^{\!4(N-1)} \right]} \left( \frac{\tilde{t}_2}{\tilde{t}_1}\right)^{\!\!3(N-1)} , \qquad &&n=1,\dots,N ;\\
        [\tilde{\mathds{H}}_{N}(\epsilon_0)]^{-1}_{2N+2n-1,2N} &= -\left( \frac{\tilde{t}_2}{\tilde{t}_1}\right)^{\!\!n-1} \frac{1}{\tilde{t}_1}\frac{\epsilon^2_0}{\left[ \, \epsilon_0^2  + \left(\frac{\kappa}{2}\right)^{\!2}  \, \left( \frac{\tilde{t}_2}{\tilde{t}_1} \right)^{\!4(N-1)} \right]} , \qquad &&n=1,\dots,N\!-\!1 .
    \end{alignedat}
\end{equation}
Similarly, for the NHSE perturbation, we have
\begin{equation}
    \begin{alignedat}{2}
        [\tilde{\mathds{H}}_{\text{NHSE}}(\epsilon_0)]^{-1}_{2n-1,2m-1} &= \left(\frac{\tilde{t}_2}{\tilde{t}_1}\right)^{\!\!m-n} \frac{1}{-\frac{\kappa}{2}+\Theta}  \,, \qquad &&n=1,\dots,N  ;\\
        [\tilde{\mathds{H}}_{\text{NHSE}}(\epsilon_0)]^{-1}_{2n,2m-1} &= -\frac{\epsilon_0}{\tilde{t}_1}\left(\frac{\tilde{t}_1}{\tilde{t}_2}\right)^{\!\!N-m-n+1} \frac{1}{-\frac{\kappa}{2}+\Theta} \, e^{-(r+s)(N-1)} , \qquad &&n=1,\dots,m\!-\!1  ;\\
        [\tilde{\mathds{H}}_{\text{NHSE}}(\epsilon_0)]^{-1}_{2n,2m-1} &= -\frac{\epsilon_0}{\tilde{t}_2}\left(\frac{\tilde{t}_1}{\tilde{t}_2}\right)^{\!\!N-m-n} \frac{1}{-\frac{\kappa}{2}+\Theta} \, e^{(r+s)(N-1)}   , \qquad &&n=m,\dots,N\!-\!1  ;\\
        [\tilde{\mathds{H}}_{\text{NHSE}}(\epsilon_0)]^{-1}_{2N+2n-2,2m-1} &=  0 , \qquad &&n=1,\dots,N ;\\
        [\tilde{\mathds{H}}_{\text{NHSE}}(\epsilon_0)]^{-1}_{2N+2n-1,2m-1} &=  0 , \qquad &&n=1,\dots,N\!-\!1 ;\\
        [\tilde{\mathds{H}}_{\text{NHSE}}(\epsilon_0)]^{-1}_{2n-1,2N+2m-2} &=  0 , \qquad &&n=1,\dots,N  ;\\
        [\tilde{\mathds{H}}_{\text{NHSE}}(\epsilon_0)]^{-1}_{2n,2N+2m-2} &=  0 , \qquad &&n=1,\dots,N\!-\!1  ;\\
        [\tilde{\mathds{H}}_{\text{NHSE}}(\epsilon_0)]^{-1}_{2N+2n-2,2N+2m-2} &= -\left(\frac{\tilde{t}_2}{\tilde{t}_1}\right)^{\!\!m-n} \frac{1}{\frac{\kappa}{2}+\Theta}  , \qquad &&n=1,\dots,N ;\\
        [\tilde{\mathds{H}}_{\text{NHSE}}(\epsilon_0)]^{-1}_{2N+2n-1,2N+2m-2} &= \frac{\epsilon_0}{\tilde{t}_1}\left(\frac{\tilde{t}_1}{\tilde{t}_2}\right)^{\!\!N-m-n+1} \frac{1}{\frac{\kappa}{2}+\Theta} \, e^{(r+s)(N-1)} , \qquad &&n=1,\dots,m\!-\!1 ; \\
        [\tilde{\mathds{H}}_{\text{NHSE}}(\epsilon_0)]^{-1}_{2N+2n-1,2N+2m-2} &= \frac{\epsilon_0}{\tilde{t}_2}\left(\frac{\tilde{t}_1}{\tilde{t}_2}\right)^{\!\!N-m-n} \frac{1}{\frac{\kappa}{2}+\Theta} \, e^{-(r+s)(N-1)}  , \qquad &&n=m,\dots,N\!-\!1 ,
    \end{alignedat}
\end{equation}
where
\begin{equation}
    \Theta \equiv 2\epsilon_0\left(\frac{\tilde{t}_1}{\tilde{t}_2} \right)^{\!\!N-2m+1} \sinh[(r\!+\!s)(N\!-\!1)] .
\end{equation}


\begin{thebibliography}{100}
\bibitem{footnote} Technically, the term dynamical matrix should refer to $i$ times the matrix formed by the blocks $h^X$ and $h^P$. Note that when the driving parameter $t_1,t_2$ are set to zero, the blocks $i \, h^X$ and $i \, h^P$ combine to produce a Hermitian dynamical matrix, as expected.
\end{thebibliography}
\end{document}